\documentclass[twocolumn,12pt]{aastex61}
\begin{document}

\title{The Keck Cosmic Web Imager Integral Field Spectrograph}

\author{Patrick Morrissey}
\affiliation{Cahill Center for Astrophysics, California Institute of Technology,
1216 East California Boulevard, Pasadena, California 91125}
\affiliation{Jet Propulsion Laboratory, California Institute of Technology, 4800 Oak Grove Drive, Pasadena, CA 91109}
\correspondingauthor{Patrick Morrissey}
\email{patrick.morrissey@jpl.nasa.gov}

\author{Matuesz Matuszewski}
\affiliation{Cahill Center for Astrophysics, California Institute of Technology,
1216 East California Boulevard, Pasadena, California 91125}

\author{D. Christopher Martin}
\affiliation{Cahill Center for Astrophysics, California Institute of Technology,
1216 East California Boulevard, Pasadena, California 91125}

\author{James D. Neill}
\affiliation{Cahill Center for Astrophysics, California Institute of Technology,
1216 East California Boulevard, Pasadena, California 91125}

\author{Harland Epps}
\affiliation{University of California Observatories/Lick Observatory, 1156 High Street, Santa Cruz, CA 95064}

\author{Jason Fucik}
\affiliation{Cahill Center for Astrophysics, California Institute of Technology,
1216 East California Boulevard, Pasadena, California 91125}

\author{Bob Weber}
\affiliation{Cahill Center for Astrophysics, California Institute of Technology,
1216 East California Boulevard, Pasadena, California 91125}

\author{Behnam Darvish}
\affiliation{Cahill Center for Astrophysics, California Institute of Technology,
1216 East California Boulevard, Pasadena, California 91125}

\author{Sean Adkins}
\affiliation{Wave$_2$ Enterprises, 77-177 Halawai Way, Kailua Kona, HI 96740}

\author{Steve Allen}
\affiliation{University of California Observatories/Lick Observatory, 1156 High Street, Santa Cruz, CA 95064}

\author{Randy Bartos}
\affiliation{Jet Propulsion Laboratory, California Institute of Technology, 4800 Oak Grove Drive, Pasadena, CA 91109}

\author{Justin Belicki}
\affiliation{Cahill Center for Astrophysics, California Institute of Technology,
1216 East California Boulevard, Pasadena, California 91125}

\author{Jerry Cabak}
\affiliation{University of California Observatories/Lick Observatory, 1156 High Street, Santa Cruz, CA 95064}

\author{Shawn Callahan}
\affiliation{Large Synoptic Survey Telescope, 950 N. Cherry St., Tucson, AZ 85721}

\author{Dave Cowley}
\affiliation{University of California Observatories/Lick Observatory, 1156 High Street, Santa Cruz, CA 95064}

\author{Marty Crabill}
\affiliation{Cahill Center for Astrophysics, California Institute of Technology,
1216 East California Boulevard, Pasadena, California 91125}

\author{Willian Deich}
\affiliation{University of California Observatories/Lick Observatory, 1156 High Street, Santa Cruz, CA 95064}

\author{Alex Delecroix}
\affiliation{Cahill Center for Astrophysics, California Institute of Technology,
1216 East California Boulevard, Pasadena, California 91125}

\author{Greg Doppman}
\affiliation{W. M. Keck Observatory, 65-1120 Mamalahoa Highway, Kamuela, HI, USA 96743}

\author{David Hilyard}
\affiliation{University of California Observatories/Lick Observatory, 1156 High Street, Santa Cruz, CA 95064}

\author{Ean James}
\affiliation{W. M. Keck Observatory, 65-1120 Mamalahoa Highway, Kamuela, HI, USA 96743}

\author{Steve Kaye}
\affiliation{Cahill Center for Astrophysics, California Institute of Technology,
1216 East California Boulevard, Pasadena, California 91125}

\author{Michael Kokorowski}
\affiliation{Jet Propulsion Laboratory, California Institute of Technology, 4800 Oak Grove Drive, Pasadena, CA 91109}

\author{Shui Kwok}
\affiliation{W. M. Keck Observatory, 65-1120 Mamalahoa Highway, Kamuela, HI, USA 96743}

\author{Kyle Lanclos}
\affiliation{W. M. Keck Observatory, 65-1120 Mamalahoa Highway, Kamuela, HI, USA 96743}

\author{Steve Milner}
\affiliation{W. M. Keck Observatory, 65-1120 Mamalahoa Highway, Kamuela, HI, USA 96743}

\author{Anna Moore}
\affiliation{Research School of Astronomy and Astrophysics, The Australian National University, Canberra, ACT, Australia}

\author{Donal O'Sullivan}
\affiliation{Cahill Center for Astrophysics, California Institute of Technology,
1216 East California Boulevard, Pasadena, California 91125}

\author{Prachi Parihar}
\affiliation{Cahill Center for Astrophysics, California Institute of Technology,
1216 East California Boulevard, Pasadena, California 91125}

\author{Sam Park}
\affiliation{W. M. Keck Observatory, 65-1120 Mamalahoa Highway, Kamuela, HI, USA 96743}

\author{Andrew Phillips}
\affiliation{University of California Observatories/Lick Observatory, 1156 High Street, Santa Cruz, CA 95064}

\author{Luca Rizzi}
\affiliation{W. M. Keck Observatory, 65-1120 Mamalahoa Highway, Kamuela, HI, USA 96743}

\author{Constance Rockosi}
\affiliation{University of California Observatories/Lick Observatory, 1156 High Street, Santa Cruz, CA 95064}

\author{Hector Rodriguez}
\affiliation{Cahill Center for Astrophysics, California Institute of Technology,
1216 East California Boulevard, Pasadena, California 91125}

\author{Yves Salaun}
\affiliation{Winlight Optics, 135 Rue Benjamin Franklin, Pertuis, France 84120}

\author{Kirk Seaman}
\affiliation{Jet Propulsion Laboratory, California Institute of Technology, 4800 Oak Grove Drive, Pasadena, CA 91109}

\author{David Sheikh}
\affiliation{ZeCoat Corporation, 23510 Telo Ave, \#3, Torrance, CA 90505}

\author{Jason Weiss}
\affiliation{University of California, Los Angeles, CA, USA 90095-1547}

\author{Ray Zarzaca}
\affiliation{Cahill Center for Astrophysics, California Institute of Technology,
1216 East California Boulevard, Pasadena, California 91125}

\received{20 February 2018}
\revised{4 July 2018}
\accepted{12 July 2018}
\submitjournal{The Astrophysical Journal}

\begin{abstract}
We report on the design and performance of the Keck Cosmic Web Imager (KCWI), a general purpose optical integral
field spectrograph that has been installed at the Nasmyth port of the 10~m Keck II telescope on Mauna Kea, HI.  
The novel design provides blue-optimized seeing-limited imaging from 350-560~nm with configurable spectral resolution from $1000 - 20000$ in a
field of view up to $20\arcsec \times 33\arcsec$.  
Selectable volume phase holographic (VPH) gratings and high
performance dielectric, multilayer silver and enhanced aluminum
coatings provide end-to-end peak efficiency in excess of 45\% while accommodating the future addition
of a red channel that will extend wavelength coverage to 1~micron.
KCWI takes full advantage of the excellent seeing and dark sky above Mauna Kea with an available
nod-and-shuffle observing mode.  The instrument is
optimized for observations of faint, diffuse objects such as the
intergalactic medium or cosmic web.  In this paper, a detailed
description of the instrument design is provided with measured
performance results from the laboratory test program and ten nights
of on-sky commissioning during the spring of 2017.  The KCWI team is
lead by Caltech and JPL (project management, design and implementation) in
partnership with the University of California at Santa Cruz (camera
optical and mechanical design) and the W.~M.~Keck Observatory (observatory interfaces).
\end{abstract}


\section{Introduction}
\label{s:intro}
In recent years, integral field spectrographs have proliferated at
major observatories around the world (\citet{allington02}; \citet{bacon01,bacon15}; \citet{larkin06}; \citet{zurlo14}).  These powerful new instruments produce three dimensional images in which each image pixel contains full spectral information. They can be optimized for a diverse array of scientific applications from galactic
surveys to exoplanet exploration but share common advantages over
conventional slit spectrographs: high efficiency, uniform sampling,
relaxed pointing, and unmatched survey capability.  These features can
be combined with a large telescope under dark skies to realize an
instrument that is ideally suited for faint, diffuse light
astronomical observations (\citet{morrissey12}).  We have conceived of such a spectrograph
designed to probe the sky to extremely faint levels ($\sim1$\% sky) with
minimal systematics by making use of reflective image slicers (\citet{bowen38}) of novel design combined with the nod-and-shuffle technique
(\citet{glazebrook01,sembach96}) to study faint extended objects such as the
circum-QSO medium (\citet{martin14a}), circumgalactic medium (\citet{rubin17}) and ultimately, the cosmic web (\citet{martin14b}) that weaves the universe together.

We are reporting on the design and performance of the Keck Cosmic Web Imager (KCWI), an integral
field spectrograph at the 10~m Keck II telescope on Mauna Kea, Hawaii that
was commissioned on sky in spring 2017.  It is a blue-sensitive
instrument (350-560~nm) with unique features, including a selectable
spectral resolution from R=1000-20000 (achieved with three image slicers of different width combined with a suite of gratings), 
a deployable nod-and-shuffle
detector mask, and a design fully compatible with the future addition
of a parallel red channel that will extend the wavelength sensitivity to 1~micron.  KCWI was devised as a general purpose
spectrograph based on a prototype that has
been operating successfully on the Palomar Hale telescope since 2009
(\citet{matmat10,martin15}).  In order to generalize the Palomar
implementation (Palomar Cosmic Web Imager, or PCWI) for facility use at Keck, we developed a set of top
level requirements to broaden its applicability to the needs of a wide
range of likely observers while providing a competitive advantage over
other available instruments in specific areas:
\begin{enumerate}
\item The instrument should provide broad optical spectral coverage,
  extending into the technically challenging blue region 350-400 nm
  while accommodating the phased implementation of a future red
  channel that would add coverage at higher redshifts.
\item
The instrument should provide high spectral resolution in order to
enable new science applications at high resolution on small scales,
such as the search for globular cluster black holes and studies of the
nuclei of active galaxies.
\item The instrument should provide moderate field survey capability,
  taking advantage of the excellent Mauna Kea site with a design that
  fully resolves the available seeing and minimizes background
  subtraction systematics
\item The instrument should be stable, easily configurable, and
  operationally straight-forward to efficiently make use of the
  available observing time.
\item The instrument should draw from the technical heritage of the
  Palomar instrument both as a proof-of-concept and as a basis for
  development of the data pipeline.
\end{enumerate}
With these in mind, we designed KCWI based on our Palomar
design, but adding elements to improve the imaging capabilities and
the range of spectral modes while still fitting into the allowed
volume envelope at the observatory.

In this paper, we will describe the instrument design and its
performance.  In Section~\ref{s:optics} we will describe the design of the optics,
camera, and detector.  In Section~\ref{s:coatings} we will describe the
optical coatings.  In Section~\ref{drp} we describe the instrument pipeline and pipeline products.  In Section~\ref{s:performance} we describe the instrument
performance measured in the laboratory and on sky during
commissioning.  In Section~\ref{s:examples} we present some representative observing
examples and finally in Section~\ref{s:discussion} we provide a summary and basis for future work..

\section{Instrument Design}
\label{s:optics}

KCWI is a highly configurable integral field spectrograph that is
optimized for seeing-limited observations of faint, diffuse light
through the use of the nod and shuffle technique in concert
with a fixed-gravity bench-mounted structure that minimizes flexure.
Its novel optical design includes selectable image slicers and
gratings that can provide spectral resolution from $1000 - 20000$ in
a field of regard $20\arcsec \times 8-34\arcsec$ (the selectable field width is inverse to spectral resolution).  Simultaneous pixel-limited
imaging and spectral resolution are achieved with the
inclusion of a cylindrical wavefront-correcting mirror, while high
throughput results from a combination of dielectric, multilayer silver
and aluminum coatings, and a blue-optimized Teledyne-e2v (Chelmsford, UK) charge coupled detector (CCD).
KCWI has been designed from the ground up as a dual channel
instrument; a 530-1050~nm channel can be added by replacing a fold
mirror with a dichroic to extend the current 350-560~nm sensitivity.

\begin{figure}
\plotone{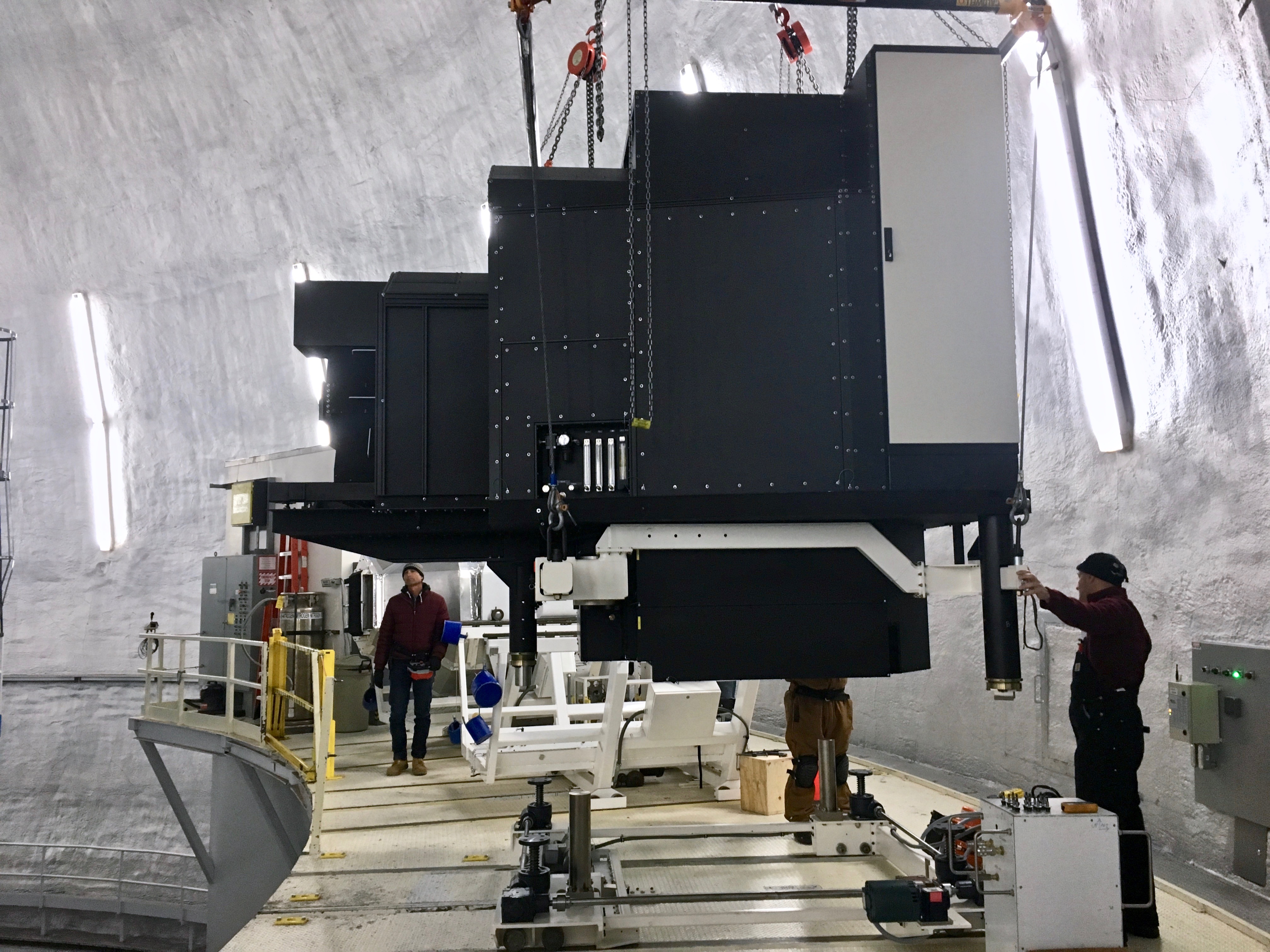}
\caption{The KCWI instrument during installation at the Keck II telescope in January 2017.\label{f:install}}
\end{figure}

KCWI operates at the right Nasmyth port of the 10~m Keck II
telescope. Figure~\ref{f:install} shows the KCWI instrument during installation at the Keck II telescope.  The instrument is one of the largest at the observatory, 
comparable to DEIMOS (\citet{faber03}) in volume ($3.0 \times 2.3 \times 3.9$~m$^3$) and weighing
approximately 4000~kg.  Major mechanisms include a K-Mirror field
de-rotator, grating exchanger and camera articulation stage allowing
completely remote configuration. A large custom made steel optical bench
(Newport Corp, Irvine, CA) supports all of the instrument's
components, and three kinematic mounting points accurately position
the instrument at the telescope focus. Transport rails on the Nasmyth
platform and deck allow the instrument to be moved between the
focal station and a storage location using an attached transport
cart.

The instrument is housed in an enclosure and purged with filtered, dry air
to protect the optical components (particularly the multi-layer silver
coatings) from humidity, dust and other contaminants. Large panels are
removable to provide service access.  The camera and gratings are
mounted on the bottom of the optical bench, while the K-Mirror, IFU,
and collimating optics are mounted on the top. The electronics are
mounted in an electrically shielded, glycol cooled equipment rack that
provides service access to all of the electronics except for the CCD
controller, which is mounted near the detector assembly.

\begin{figure*}
\plotone{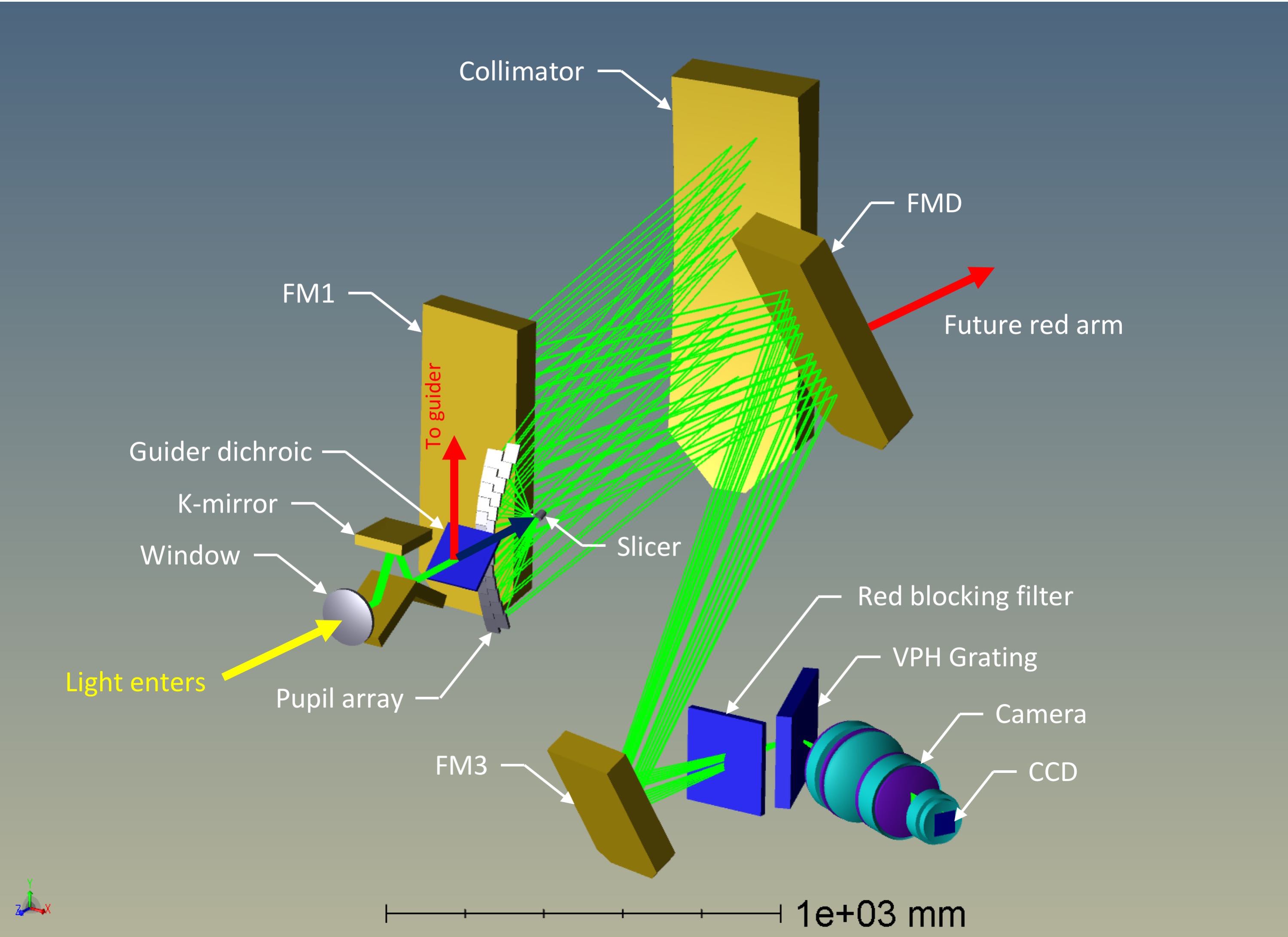}
\caption{The optical layout of KCWI showing a subset of the optical rays for clarity.  The primary mirror is 1.1~m tall for scale.  The system forms a hexagonal compound pupil image at the grating that is 160 mm in diameter simultaneously with pixel-limited re-imaging of the slicer focal plane at the detector.  A parallel red arm can be accommodated by replacing the FMD fold mirror with a dichroic.\label{f:kcwi3d}}
\end{figure*}

The optical layout of KCWI is illustrated in Figure~\ref{f:kcwi3d}, with
key design parameters summarized in Table~\ref{t:kcwiparams}.  Converging light from the telescope tertiary (f/13.7 measured across the vertices of the hexagonal Keck primary) is directed through the instrument
protective hatch, which is a $145 \times 135$~mm twin-blade shutter made by Bonn-Shutter UG (Bonn, Germany).  The shutter is a spare identical to the one used in the camera, and achieves linearity better than 1\% for exposures as short as 0.1~s (c.f. \citet{reif99}).  Light continues through an AR-coated, fused silica meniscus
entrance window that is designed to operate at its aplanatic conjugate, preserving the quality of the incoming converging beam (c.f. \citet{morrissey94}) while minimizing ghosting at the focal plane.  

\begin{deluxetable*}{rcl} 
\tablecaption{The KCWI Optical Prescription\label{t:kcwiparams}}
\tablehead{
\colhead{Parameter} & \colhead{Value} & \colhead{Notes}
}
\startdata
Design band                 & 350-560 nm & Blue channel\\
Design focal ratio          & 13.7       & Keck telescope\\
Telescope Plate Scale       & 1.375\arcsec-mm$^{-1}$\\
Field of View               & (8.25, 16.5, or 33)$\arcsec \times 20\arcsec$ & Selectable\\
Slit width                  & (0.34, 0.69, or 1.38)$\arcsec \times 24$      & Selectable\\
Window front radius         & 391.7 mm   & Convex, spherical\\
Window rear radius          & 382.6 mm   & Concave, spherical\\
IFU radius                  & 225 mm     & Focus to vertex\\
Slice radius                & 5206 mm    & Convex, spherical\\
Pupil mirror radius         & Infinity   & \\
Collimator radius           & 4345 mm    & Concave, spherical\\
FM1 radius\tablenotemark{a} & 675 m      & Convex, cylindrical\\
Beam diameter               & 160 mm     & Compound pupil\\
Camera focal length         & 304.4 mm   & All spherical\\
Science detector\tablenotemark{b}   & $4096\times 4112$  & 15 micron pixels\\
Science detector plate scale        & 0.147\arcsec-pixel$^{-1}$ & 1x1 bins\\
Focal plane camera          & $3296\times 2472$ & $24.9\arcsec \times 18.7\arcsec$\\
Guider (Magiq) camera       & $1024\times 1024$ & $3.1' \times 3.1'$\\
\enddata
\tablenotetext{a}{The FM1 cylinder axis is vertical, along the long axis of the mirror.}
\tablenotetext{b}{The KCWI detector is a standard Teledyne-e2v CCD231-84 thin, back-illuminated, Astro-BB process CCD.}
\end{deluxetable*}

\begin{figure*}
\plotone{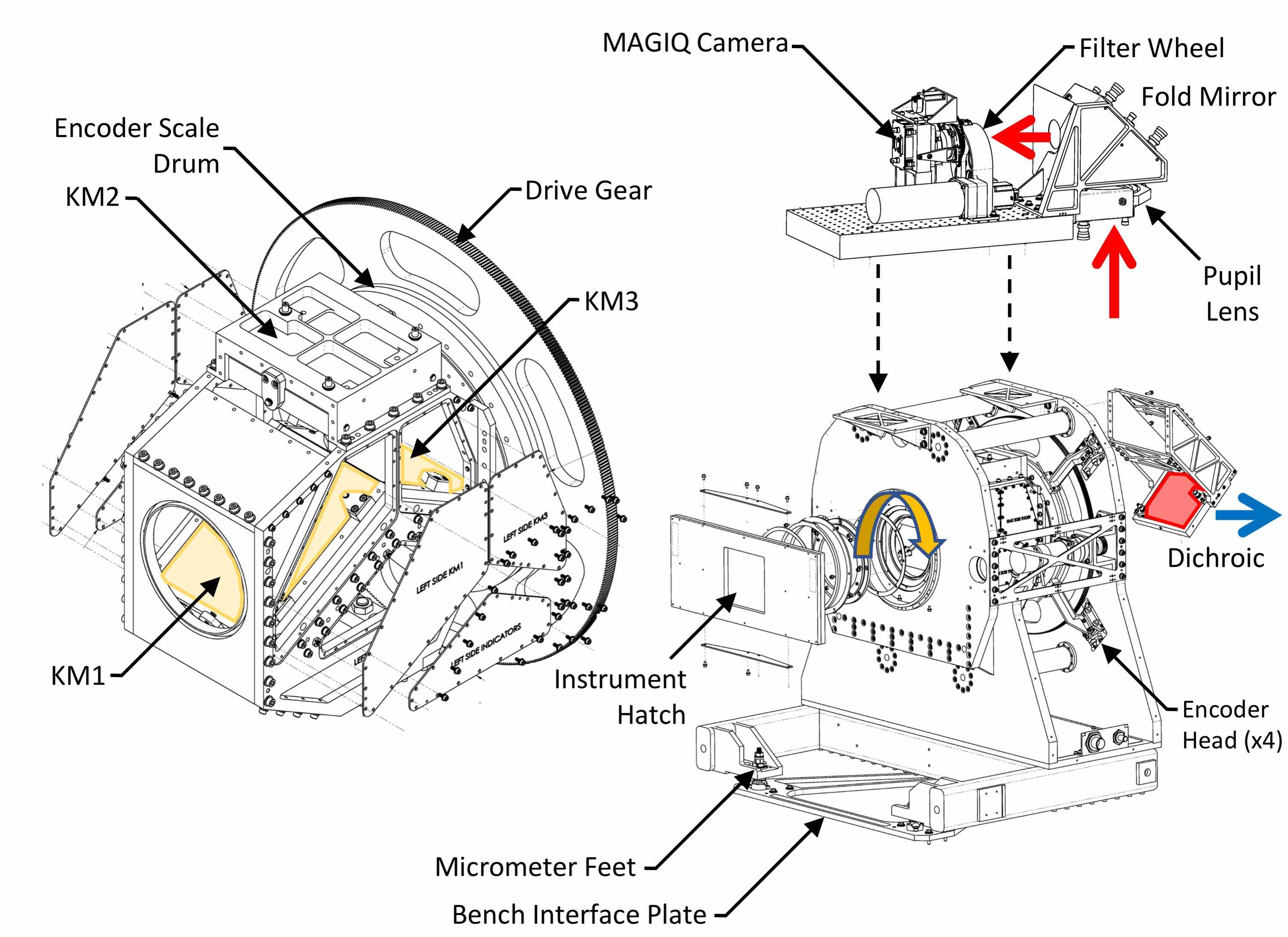}
\caption{A line drawing of the K-mirror assembly, including the instrument hatch, guider dichroic, and MAGIQ guider.  On the left, an expanded view of the rotating turret assembly is shown.  Mirrors are kinematically mounted inside the turret and can be adjusted with locking spring plungers that rest on silicon carbide pucks bonded to each mirror surface.  On the right, the turret is shown inside the K-mirror assembly.  The assembly also supports the guider dichroic, which directs red light to a MAGIQ guider that is bench mounted onto the top of the structure.\label{f:kmassy}}
\end{figure*}

\begin{figure}
\plotone{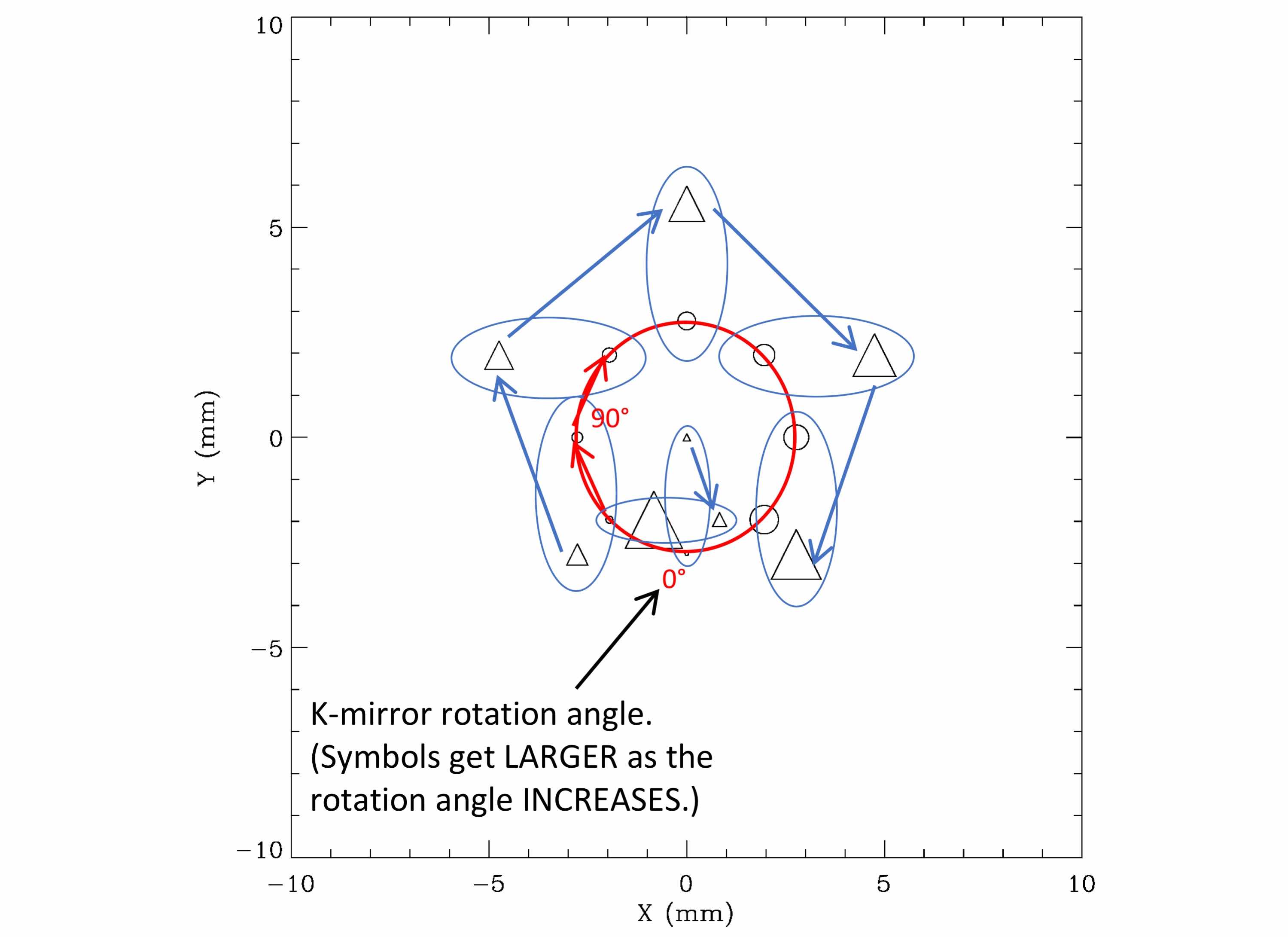}
\caption{An example of focal plane image motions that result from a combination of internal and external K-mirror misalignments.  The circular symbols show the path of a ``planet'' on the optical axis of the telescope but deflected by a misaligned K-mirror.  This planet executes a circle (depicted in red) in the focal plane at the same rate as the K-mirror.  The triangular symbols show the path of a ``moon'' that is off the optical axis of the telescope.  This triangular moon ``orbits'' the planet at twice the rate of the K-mirror.  At a K-Mirror rotation of 180$^{\circ}$, the pattern of ``moon'' and ''planet'' has rotated 360$^{\circ}$, but is displaced in the focal plane.  In this figure, the size of the symbol is proportional to the K-mirror rotation angle.\label{f:kmwobble}}
\end{figure}

The converging beam continues through a K-mirror image de-rotator to deliver a stationary, on-axis, un-vignetted 40\arcsec\ diameter science field to a selectable image slicer located at the telescope focus.  The
de-rotator contains three broad band
(350-1050~nm) dielectric coated Zerodur (Schott AG, Germany) mirrors each with an average
reflectivity of over 99\%.  A diagram of the K-mirror assembly is shown in Figure~\ref{f:kmassy}.  Its structure is comprised of a pair of weight-relieved steel plates bridged by cylindrical beams and kinematically mounted through three micrometer feet to a steel interface plate, which is in turn bolted to the optical bench.  The mechanism is a gear driven turret structure that is cantilevered off of the rear plate.  The turret, which contains the three mirrors, is supported by a turntable bearing. Its motion is monitored by a four-head encoder and optical drum scale read by a Galil 4010 controller with custom firmware.  The K-mirror has the property that it rotates the telescope field at twice the speed of its own rotation.
An example is shown in Figure~\ref{f:kmwobble}.  A star on the rotational axis of an internally misaligned K-mirror will execute circular motions in the focal plane at the same speed as the mechanism, while a star off the axis (externally misaligned) of a correctly aligned K-mirror will execute circular rotations in the focal plane at {\em twice} the rate of the mechanism.  Combinations of the two misalignments can produce complex nested circular image motions. Of chief concern in the KCWI implementation is not the image stability at the focal plane, which is corrected in real time by the downstream guider, but rather the residual pointing of the beam inside the instrument, or ``beam walk'' which could result in vignetting inside the IFU pupil array.\footnote{Clearances in the pupil array are as small as 125 microns between the beam footprints and their associated mirror edges.  Vignetting in these areas would primarily affect the edges of the field of view.}  In order to ensure that the KCWI K-mirror would be aligned with sufficient precision to avoid this ($\lesssim 30\arcsec$), a Monte Carlo simulation of optical ray traces was used to validate the alignment procedure.  We determined that beam walk could be reduced to an acceptable level by aligning the three mirrors initially to FARO-arm (Lake Mary, FL) tolerance ($\lesssim 100$ microns) and then using an On-Trak (Irvine, CA) laser alignment tool to null the image motion with final adjustments of only the KM2 mirror.  

A 160 mm square dichroic beamsplitter (Asahi Spectra Co., Japan) with a 630~nm cut-off wavelength
is mounted between the rear of the K-Mirror and the focal plane.  This
element reflects red light to a standard platform-mounted MAGIQ guider (\citet{adkins08}) with approximately a 2\arcmin\ diameter field of view, while blue light is transmitted to the science
instrument.  The coating performance is described in Section~\ref{s:coatings}.  This element would be removed to accommodate the addition of a red channel.  In the current implementation, the downstream guider location enabled by the dichroic permits measurements of the seasonal performance of the de-rotator necessary to inform any possible change to a guider location upstream of the K-Mirror as part of the red channel implementation.

\begin{figure}
\plotone{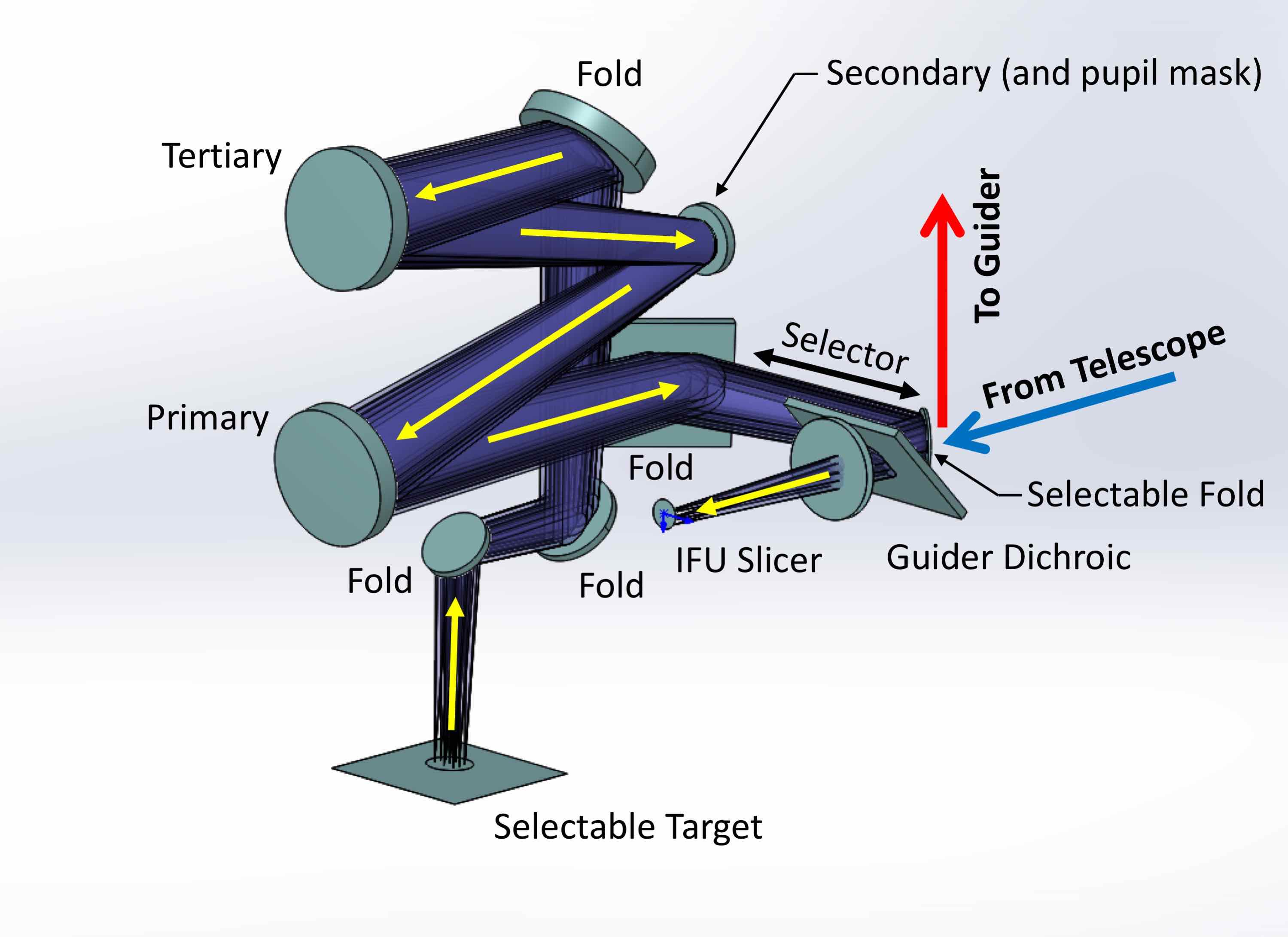}
\caption{The optical layout of the KCWI Offner relay calibration system.  The system re-images a target at 1:1 magnification onto the image slicer through the use of a selectable periscope mirror.  The system can simultaneously illuminate the guider and the spectrograph to aid alignment.  The relay also simulates a hexagonal telescope pupil with a rotatable mask at the secondary mirror that is correctly relayed onto the grating.  The selectable calibration target is illuminated through an integrating sphere by either an LED continuum lamp, or by ThAr or FeAr hollow cathode arc lamps.\label{f:calunit}}
\end{figure}

The instrument includes a calibration subsystem as shown in Figure~\ref{f:calunit}.  It
utilizes a pair of deployable periscopes to inject light through (or
after) the guider dichroic to the guider and the science focal
planes.  The unit employs a reflective Offner relay that re-images a
set of moveable targets at 1:1 magnification onto the science focal
plane.  The relay also simulates the hexagonal pupil of the telescope
primary with a rotatable mask mounted over the relay secondary mirror.
The calibration targets include an
array of pinholes, bars and flat field elements.  The
calibration subsystem is illuminated through an integrating sphere by
a selectable set of lamps including an LED continuum source, as well
as ThAr and FeAr hollow cathode lamps.

The telescope focal
plane is divided into 24 slitlets by one of three selectable image
slicer elements of the KCWI Integral Field Unit
(IFU), which was fabricated by Winlight Optics (Pertuis, France).  The slicer mirror stacks provide 0.35\arcsec, 0.7\arcsec\ or 1.4\arcsec\
spatial resolution (with corresponding 8.4\arcsec, 16.8\arcsec, or 33.6\arcsec\ FOV in
one axis and a 20\arcsec\ FOV for all configurations in the other axis). 

\begin{figure}
\plotone{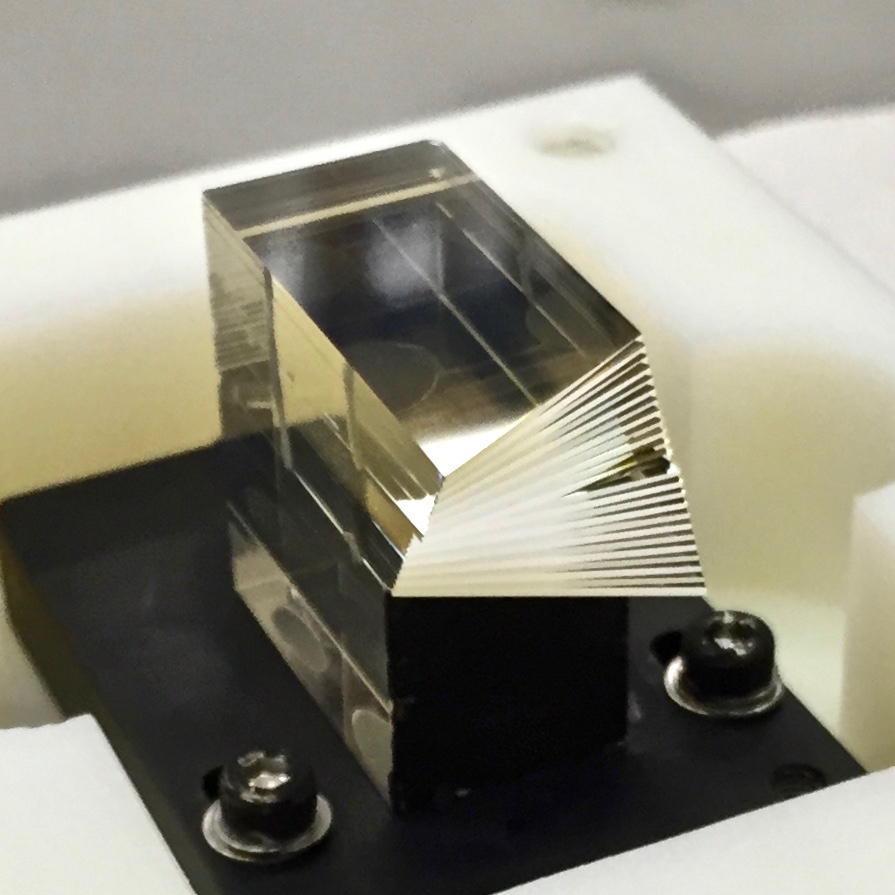}
\caption{A close-up view of the medium slicer.  As shown here the element is on its side relative to its orientation inside the instrument.  This element is made from optically contacted Zerodur.  Each of the 24 slitlets is 0.5~mm thick and 14.8~mm tall.  The reflective faces are convex with a radius of curvature of 5206~mm.  The coating is multi-layer enhanced silver.\label{f:medslc}}
\end{figure}

\begin{figure*}
\plotone{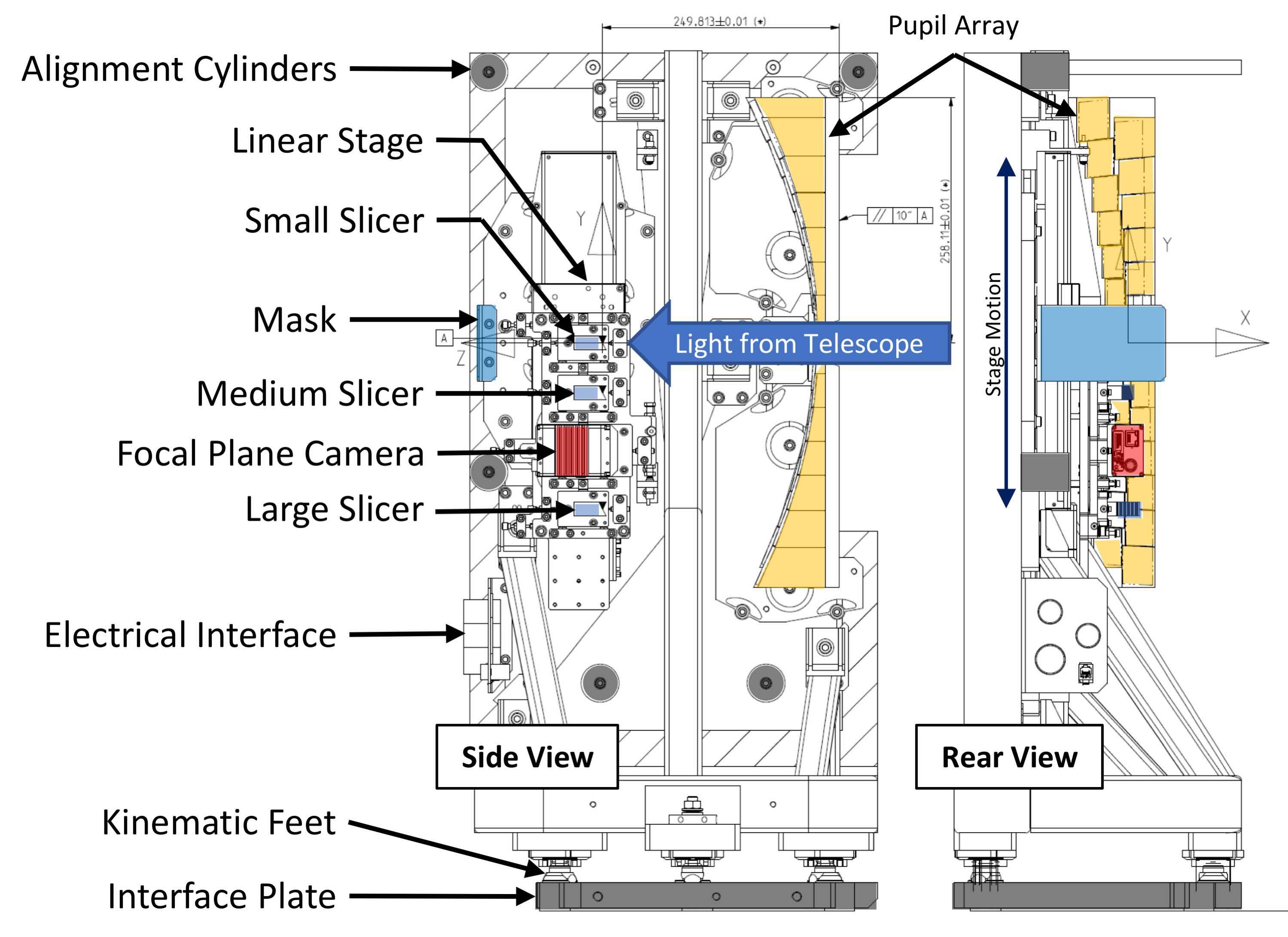}
\caption{Side and rear views of the KCWI opto-mechanical integral field unit (IFU) assembly.  One of three image slicers or a direct imaging camera can be placed in the optical path by a linear stepper motor.  The slicer elements are reconfigured in reflection behind the pupil array into a long virtual slit that is re-imaged and dispersed by the spectrograph.\label{f:ifu}}
\end{figure*}

A close-up view of the medium slicer is shown in Figure~\ref{f:medslc}.  A mechanical diagram of the IFU subassembly is shown in Figure~\ref{f:ifu}.  The image slicers sit on a linear stage that can position one of the three slicers or a focal plane camera (an Allied Vision Technologies Pro Silica GT3300 monochrome camera with an ON Semiconductor KAI-08050 CCD sensor) onto the spectrograph optical axis.  The IFU structure is comprised of a Newport optical bench with Invar face sheets and an aluminum core flexure mounted to a stainless steel weldment in such a way as to mimic a steel construction that thermally matches the K-mirror assembly.  

The Zerodur slicers work in concert with a fixed array of 24 flat ``pupil'' mirrors to reconfigure the focal plane in reflection into a vertical long slit referred to as the ``virtual slit'' (\citet{weitzel96}).  The reformatted slicer array reflection, rather than being perfectly vertical, assumes a staircase pattern resulting from the geometry of the pupil array that
is required to avoid overlap of the diverging beams of light from the
slicer elements.  The virtual slit pattern has also been tuned (lateral curvature added) to cancel long slit curvature in images taken with the high resolution gratings,
providing maximum bandwidth on the CCD (limited only by the staircase).  This tuning was achieved by weighting the optical merit function to control the position of slice images on the detector in the high spectral resolution configuration.  A consequence of this tuning is to add a ``reverse'' slit curvature in low resolution configurations, although this was deemed to be acceptable given the larger bandwidth.

\begin{figure*}
\plotone{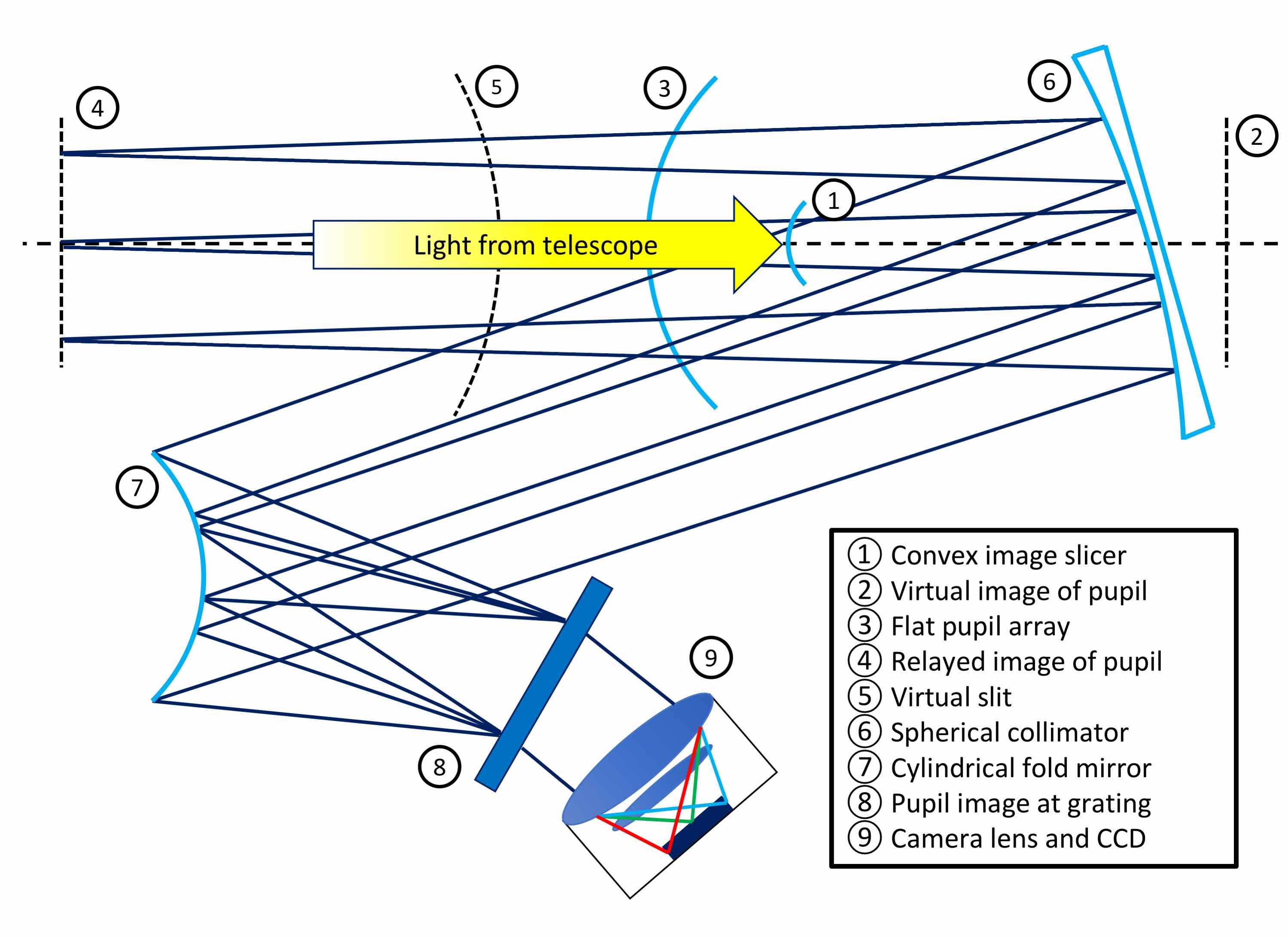}
\caption{The KCWI pupil relay system, which images the telescope primary mirror onto the instrument grating surface in parallel with the spectrograph slit relay, which images the slicer elements onto the detector.  \textit{1:} Light from the telescope arrives at the convex reflective slicer. \textit{2:} The virtual image of the telescope pupil is located behind the slicer. \textit{3:} The array of 24 flat pupil mirrors.  \textit{4:} The virtual image of the telescope pupil as relayed by the pupil mirrors. \textit{5:} The virtual slit is formed by the pupil mirrors and located behind the pupil array. \textit{6:} The spherical spectrograph collimator. \textit{7:} The wavefront-correcting cylindrical fold mirror.  \textit{8:} The compound pupil image is formed at the grating, which is located at the radius of the collimator. \textit{9:} The camera lens forms an image of the virtual slit on the detector.
\label{f:puprelay}}
\end{figure*}

\begin{figure*}
\plotone{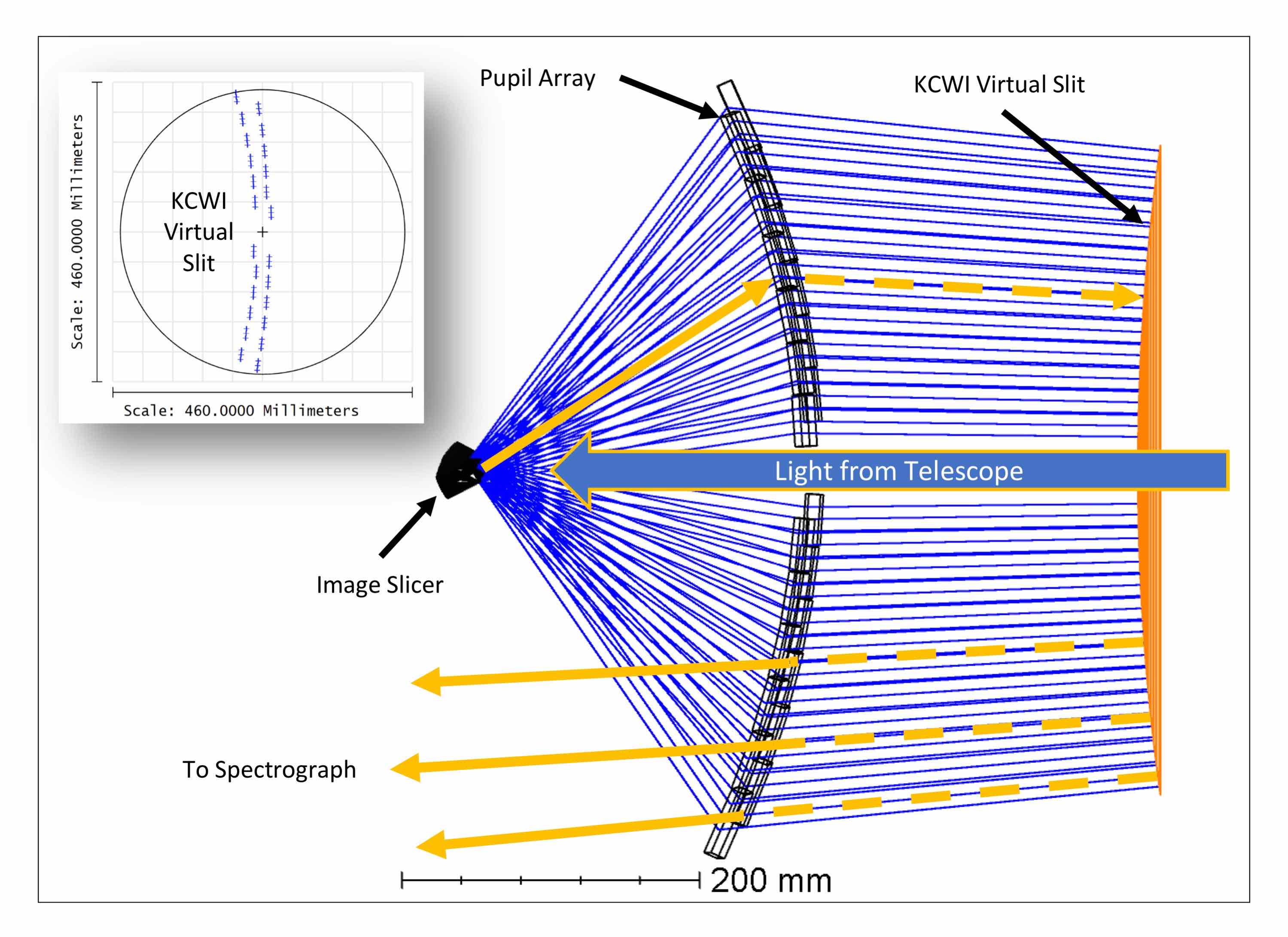}
\caption{A ray trace of the KCWI IFU and virtual slit.  Light enters from the right and is focused on the slicer at the telescope focal plane.  Individual slices are tilted to re-direct light towards the pupil array, which in turn forms the virtual slit (shown in orange) that is re-imaged by the spectrograph onto the detector.  The inset shows the footprint of the virtual slit on the spherical virtual slit surface.\label{f:vslit}}
\end{figure*}

The individual slices are each spherically convex in order to re-image the
telescope pupil onto the grating.  Each slice acts as a field lens to form the pupil image, while the IFU works in concert to ensure the 24 individual pupil images are simultaneously co-aligned on the grating to form a compound pupil image.  This configuration ensures that the minimum area of the grating-camera system is used, maximizing image quality while minimizing the required grating and camera sizes.  The pupil relay system is illustrated in Figure~\ref{f:puprelay}.  It can be interpreted as a Schmidt camera in reverse, with the ``entrance'' at the location of the compound pupil on the grating, which is formed at the radius of curvature of the spectrograph collimator.  On the other side of the relay, the IFU virtual slit, which is curved reminiscent of a Schmidt camera focal plane, is formed at the collimator focal point.  The virtual slit curvature is three dimensional as shown in Figure~\ref{f:vslit}: a ``C'' shaped track of twenty four slitlets along the surface of a sphere roughly equal in radius to the collimator focal length.  The diverging light from the virtual slit proceeds to
a spherical collimator, and then to a series of three fold mirrors.

\begin{figure*}
\plotone{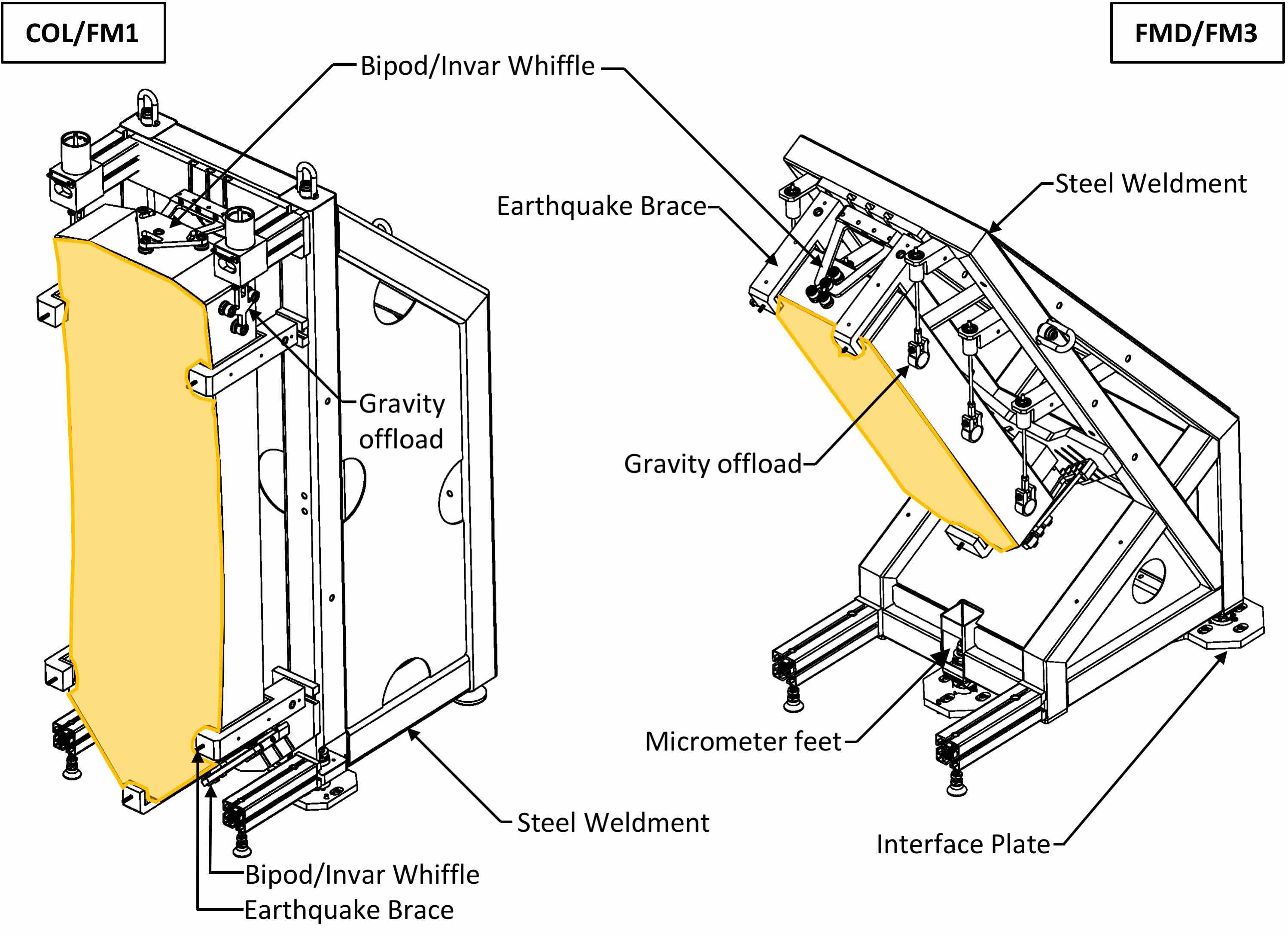}
\caption{Line drawings of the two styles of KCWI large optics mounts.  On the left is the design for the vertical mirrors, the COL and FM1 (similar but not identical).  A tubular steel weldment structure provides stiffness and is CTE-matched to the K-mirror structure.  The weldment is aligned with three micrometer feet and attached to the steel bench with a steel interface plate.  In this design, the Zerodur mirror hangs from the weldment by a pair of suspension springs that are bonded to the mirror with Invar pads.  The mirror is then held in position kinematically by three bipod-mounted whiffles, each with three Invar bond pads.  The whiffles and pads are attached by spherical bearings.  On the right the design for the FMD relay mirror is shown.  The FM3 mirror is conceptually similar (and is at the same angle) but is mounted from above rather than below.  The tilted FMD and FM3 mounts use the same principle of offloading the mirror weight on suspension springs while constraining the position kinematically with three bipod mounted whiffles.  Each of the four mirrors are held in place by 15 bond pads, providing risk reduction in case of a bond failure during handling.\label{f:lomounts}}
\end{figure*}

The large rectangular Zerodur optical elements range in height from 1.1~m (COL) to 400~mm (FM3), decreasing as the collimated virtual slit converges to the compound pupil image located at the grating.  The mirrors are 
each bonded to a set of three 3-point semi-kinematic
Invar whiffle trees (nine bonded locations) that are flexure mounted to tubular steel weldments as
shown in Figure~\ref{f:lomounts}.  There are two similar optical mount designs to
accommodate vertical (COL, FM1) and tilted (FMD, FM3) mirrors.  Both designs provide weight
relief to the heavy mirrors through sets of bonded gravity offloading
suspension springs.  One can think of the mirrors suspended from the springs, and then clamped into position by a kinematic interface.  Prior to bonding, the Zerodur and Invar surfaces were abraded with 220~grit silicon carbide and water until uniform in appearance.   The parts were rinsed in water and flushed with acetone and alcohol prior to being coated with Dow Corning (Auburn, MI) Z-6020 Silane primer (0.5\% solution in ethanol).  Approximately 30 minutes after priming, the parts were bonded with Armstrong A12 epoxy mixed 1.5:1 with 1\% 180~micron glass beads added by weight to set the bond line.  Squeeze-out was carefully removed to eliminate the possibility of high-CTE epoxy fracturing the glass under thermal stress.  Bonded parts were clamped in place for 7 days at room temperature to cure.  This process was qualified to hold the 105~kg collimator mirror under a 4g load through a temperature range from -20 to +35~$^{\circ}$C.  

The FM1 fold mirror is a 0.675~km radius convex
cylinder (axis along the height of the mirror) providing wavefront correction to compensate the collimator
working angle, while FMD and FM3 are flat.  The FMD fold mirror relays the light below the bench and is
designed to be replaced with a dichroic to accommodate a future parallel red
channel.

On the lower side of the bench, the FM3 fold mirror
redirects the beam through a 248x220~mm red light blocking dielectric
filter (Asahi Spectra Co., Tokyo, Japan) with a 570 nm 50\% cutoff. The substrate is 20 mm thick Corning 7980 fused silica, tilted at an angle of 15$^{\circ}$ to the incoming beam to reduce ghost light between the detector and filter.  The transmitted blue light forms a 160~mm diameter compound pupil image at the selected grating surface.

\subsection{Gratings}

KCWI accommodates a suite of five Volume Phase Holographic (VPH) transmission gratings that are designed to maximize the flexibility and efficiency of the instrument.  These gratings are three dimensional structures that simultaneously obey the classical grating equation and exhibit a Bragg efficiency peak (\citet{barden98, barden00}). The efficiency at a given wavelength can be optimized by rotating the grating between the collimator and camera.  Design advantages include a reduction in ghosts and scattered light typically associated with surface relief gratings, and complete encapsulation between two protective glass covers that are AR coated to increase performance.  The ability to tune the fringes within the volume of the grating allows the designer to optimize the polarization efficiency (\citet{baldry04}).  The VPH grating can be seen as combining the best features of both high efficiency ruled gratings and low scatter, polarization-insensitive holographic gratings.

Our suite of five gratings was purchased from Wasatch Photonics (Logan, UT) and is shown in Table~\ref{t:gratings} with typical incidence angles and spectral ranges across the CCD.  The spectral range is reduced to about $1/3$ the value shown when the N\&S mask is deployed.  The gratings are of the ``Dickson'' design (\citet{dickson03}) that achieves simultaneous maximum efficiency for both {\em s-} and {\em p-} polarizations.  The grating gelatin is sandwiched between a pair of $350\times 250\times 20$~mm thick, AR-coated Corning 7980 (fused silica) substrates ground and polished by Sydor Optical (Rochester, NY).  The gratings are each designed to meet the instrument resolution requirements and when paired with the available image slicers provide spectral resolution in the range from 1000-20000 as shown in Table~\ref{t:gres}.  A ``fringe tilt'' of approximately $2^{\circ}$ built into the design of each effectively places the grating at a small angle relative to the glass substrates (on the axis of dispersion) so that the Bragg angle of peak efficiency redirects a bright ``Littrow'' recombination ghost --- a retro-reflection from the CCD to the grating and back (\citet{burgh07}) --- out of the wavelength range of interest, or at a minimum outside the area used for sensitive nod and shuffle observations.  This tilt introduces an asymmetry in the incidence and diffracted angles of about $9^{\circ}$. 

\begin{figure}
\plotone{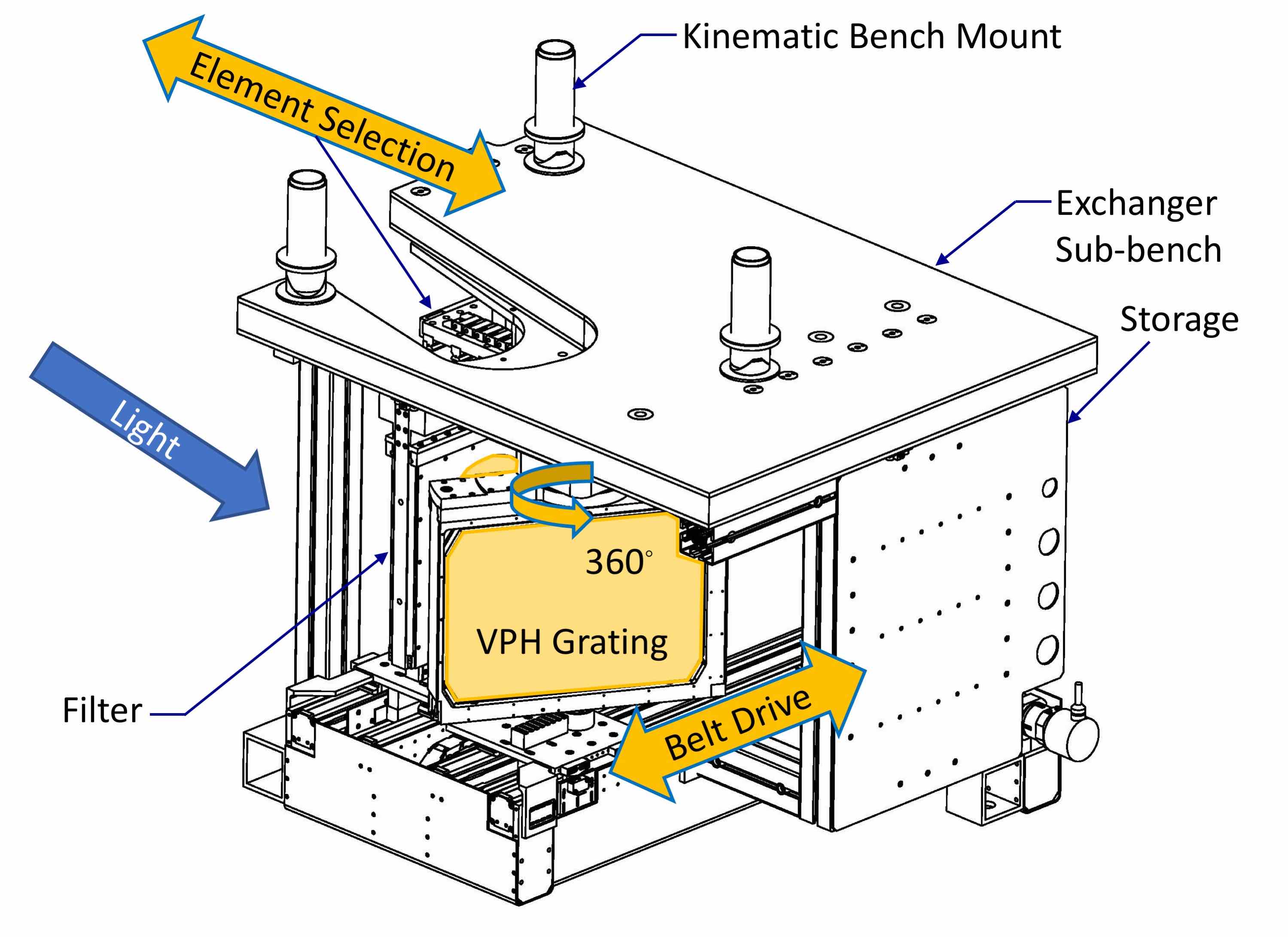}
\caption{The KCWI blue filter and grating exchanger accommodates 5 gratings and 5 filters in a jukebox mechanism.  Each element is flexure mounted into an aluminum cell and covered on each side with a sheet metal baffle appropriate for its typical operating angle.  A pair of belt driven selectors allows any grating and filter combination to be inserted into the beam.  The filter is placed at a fixed 15$^{\circ}$ angle to the beam, while the grating is rotatable to any angle.  Coordinated motions of the exchanger are executed by a Galil motion controller which completes each sequence at a home position to minimize flexure.\label{f:bex}}
\end{figure}

Each KCWI grating is flexure mounted into an aluminum cell that can be commanded in and out of an exchanger mechanism, which is shown in Figure~\ref{f:bex}.  The commanded grating can be rotated 360$^{\circ}$ without interfering with the camera at any angle.  Each grating has an oval-shaped pupil mask on either side of the cell that baffles stray light from the instrument.

\begin{deluxetable*}{cccccc}
\tablecaption{The KCWI Suite of Gratings\label{t:gratings}.}
\tablehead{
\colhead{Grating} &
\colhead{Band} &
\colhead{Rulings} &
\colhead{$\alpha$\tablenotemark{\dag}} &
\colhead{Dispersion} &
\colhead{Range\tablenotemark{\ddag}}\\
\colhead{} &
\colhead{(\AA)} &
\colhead{(lp-mm$^{-1}$)} &
\colhead{(${^\circ}$)} &
\colhead{(\AA-mm$^{-1}$)} &
\colhead{(\AA)}
}
\startdata
BH1\tablenotemark{*}  & 3500-4100 & 3800 & 55  & 6.0  &  360 \\
BH2                   & 4000-4800 & 3281 & 52  & 7.4  &  443 \\
BH3                   & 4700-5600 & 2801 & 50  & 8.6  &  516 \\
BM                    & 3500-5600 & 1900 & 30  & 16.1 &  965 \\
BL                    & 3500-5600 &  870 & 15  & 37.5 & 2250 \\
\enddata
\tablenotetext{\dag}{The typical incidence angle is shown for each grating design.  Since the gratings design is tilted, the typical dispersion angle $\beta$ is $9^{\circ}$ less than $\alpha$ at the central wavelength.}
\tablenotetext{\ddag}{The range shown is for the entire CCD and will be reduced to approximately 1/3 the value shown if the N\&S mask is deployed.}
\tablenotetext{*}{The BH1 grating is in fabrication as of this writing.  The example shown here is representative.}
\end{deluxetable*}

\begin{deluxetable}{cccc} 
\tablecaption{KCWI Slicer-Grating Spectral Resolution Table\label{t:gres}}
\tablehead{
\colhead{} & \multicolumn{3}{c}{Slicers}  \\\cline{2-4}
\colhead{Grating} & \colhead{Small} & \colhead{Medium} & \colhead{Large}\\
\colhead{} &\colhead{8.25\arcsec x 20\arcsec}& \colhead{16.5\arcsec x 20\arcsec}& \colhead{33\arcsec x 20\arcsec} 
}
\startdata
BL    &  5000 &  2500 & 1250 \\
BM    & 10000 &  5000 & 2500 \\
BHx   & 20000 & 10000 & 5000 \\
\enddata
\end{deluxetable}


\subsection{Camera}

\begin{figure}
\plotone{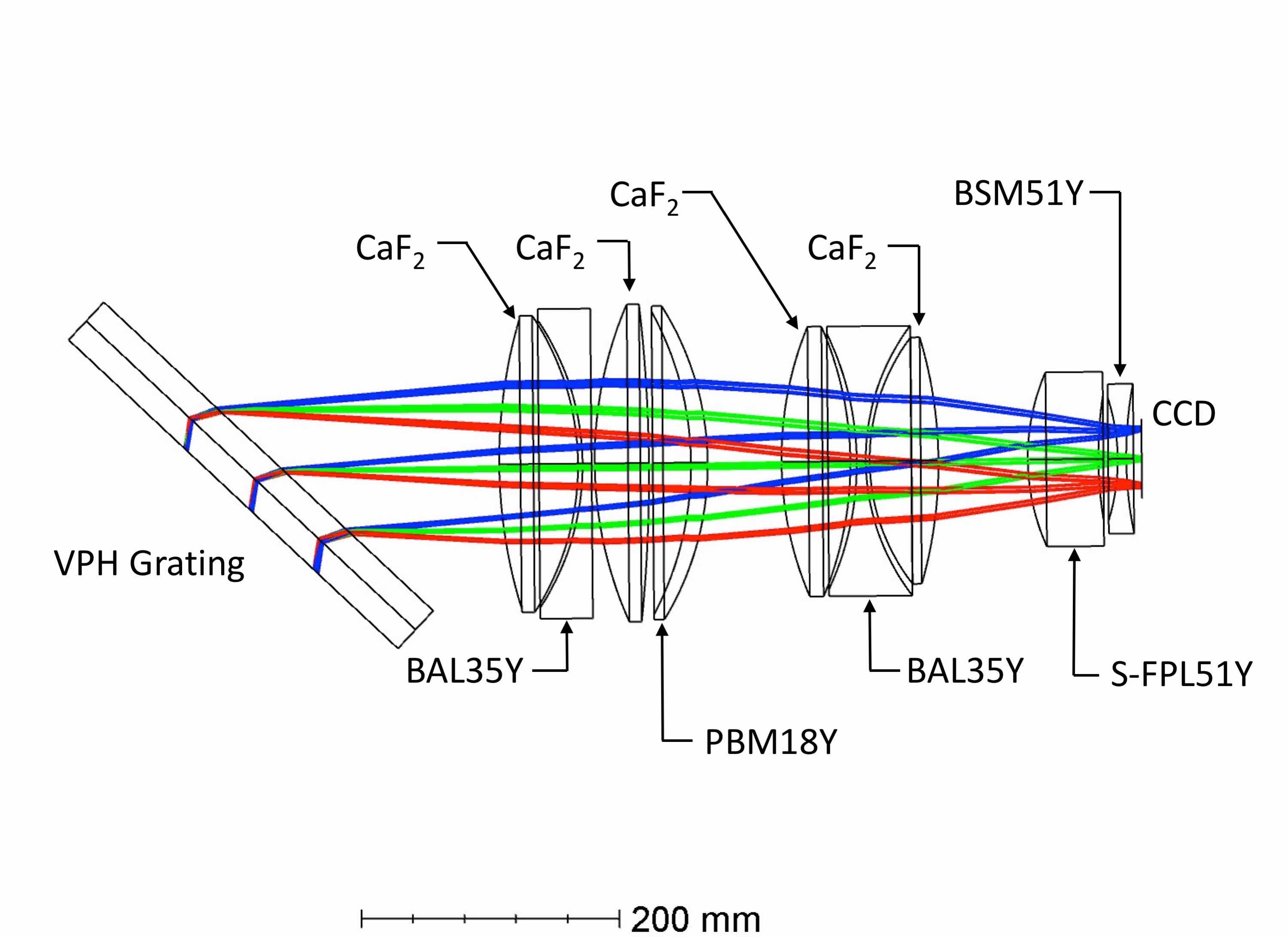}
\caption{A ray trace of the KCWI camera with high resolution grating.  The nine spherical elements were designed, fabricated and assembled at the University of California Lick Observatory in Santa Cruz.  The ninth element (BSM51Y) is a field flattener and also the vacuum window for the CCD detector.  The window and CCD move together on a linear stage to provide focus capability.\label{f:camera}}
\end{figure}

Dispersed light from the grating is re-imaged with a 304.4~mm focal length, 250~mm diameter all-spherical refractive camera designed by Harland Epps and fabricated by the team at the University of California Observatories (UCO) and Lick Observatory led by Connie Rockosi (\citet{rockosi16}).  A raytrace of the nine-element air spaced camera optics behind a high resolution grating is shown in Figure~\ref{f:camera}.  The optical prescription for typical conditions at the summit of Mauna Kea is provided in Table~\ref{t:camera} with the melt-sheet-derived Schott coefficients for the material indices of refraction provided in Table~\ref{t:melt}.  

The quartet, triplet, doublet design shown produces pixel-limited imaging over the full 350 to 560~nm design wavelength range and across an $8.2^{\circ}$ field radius (the CCD diagonal) without refocus, with acceptable
performance to 700 nm in cases where the
red blocking filter is removed at the expense of increased ghosting.  Lens materials were selected to provide high UV transmission.  Four Calcium Fluoride elements generate most of the positive optical power, while Ohara (Kanagawa, Japan) i-line glasses are used for color correction.  The last element, made of low radioactivity Ohara BSM51Y, serves as both a field flattener and detector vacuum window.  The design performance of this camera illuminated by a 158~mm diameter polychromatic pupil results in an average root-mean-square (RMS) image diameter without refocus of $15.0 \pm 4.4 \mu$m. The maximum RMS lateral color in direct imaging is $7.1 \mu$m, and distortion at the full diagonal field is 0.36\%.

\begin{deluxetable*}{rrrrrr} 
\tablecaption{KCWI Spherical Camera Prescription\tablenotemark{\dag}\label{t:camera}}
\tablehead{
\colhead{Element} & \colhead{Material} & \colhead{R$_1$} & \colhead{R$_2$} & \colhead{Thickness} & \colhead{Spacing}\\
\colhead{}        & \colhead{}         & \colhead{(mm)}  & \colhead{(mm)}  & \colhead{(mm)}      & \colhead{(mm)}
}
\startdata
1                 & CaF$_2$            &  389.859      &  -208.667     & 61.065              &  3.500 \\
2                 & BAL35Y             & -200.674      &  $\infty$     &  6.969              &  2.500 \\
3                 & CaF$_2$            &  306.107      & -1145.486     & 42.023              & 32.649 \\
4                 & PBM18Y             & -247.324      &  -230.463     & 12.947              & 57.227 \\
5                 & CaF$_2$            &  270.853      &  -263.636     & 53.418              &  5.000 \\
6                 & BAL35Y             & -230.090      &   172.635     &  6.999              &  3.000 \\
7                 & CaF$_2$            &  156.636      &  -340.233     & 53.328              & 69.390 \\
8                 & S-FPL51Y           &  167.551      &   311.824     & 54.784              & 15.000 \\
9                 & BSM51Y             & -155.177      &   300.975     &  6.589              & 11.995 \\
\enddata
\tablenotetext{\dag}{The camera is designed to operate at 2~$^{\circ}$C and 456 mmHg and to accommodate a round 160~mm diameter entrance 
pupil located 200~mm ahead of the first lens element vertex.}
\end{deluxetable*}

\begin{deluxetable*}{rrrrrrr} 
\tabletypesize{\small}
\tablecaption{Schott Coefficients for KCWI Camera Optical Glass\tablenotemark{\dag} \label{t:melt}}
\tablehead{
\colhead{Material} & \colhead{a$_0$} & \colhead{a$_1$} & \colhead{a$_2$} & \colhead{a$_3$} & \colhead{a$_4$} & \colhead{a$_5$}}
\startdata
BAL35Y   & 2.4882237 & -1.0433842$\times 10^{-2}$ & 1.3666718$\times 10^{-2}$ & 1.8847446$\times 10^{-4}$ &  9.7167343$\times 10^{-7}$ & 1.5430999$\times 10^{-7}$\\
BSM51Y   & 2.5315177 & -1.0613914$\times 10^{-2}$ & 1.4233791$\times 10^{-2}$ & 2.0023854$\times 10^{-4}$ &  1.3748829$\times 10^{-6}$ & 1.4992248$\times 10^{-7}$\\
CaF$_2$  & 2.0397792 & -3.2198575$\times 10^{-3}$ & 6.1611382$\times 10^{-3}$ & 5.2037190$\times 10^{-5}$ &  4.2463262$\times 10^{-7}$ & 7.7562224$\times 10^{-9}$\\
PBM18Y   & 2.4854756 & -9.2258527$\times 10^{-3}$ & 1.9811993$\times 10^{-2}$ & 8.3005994$\times 10^{-4}$ & -3.5512051$\times 10^{-5}$ & 4.3642513$\times 10^{-6}$\\
S-FPL51Y & 2.2192616 & -5.2736411$\times 10^{-3}$ & 8.4898019$\times 10^{-3}$ & 8.3761331$\times 10^{-5}$ &  6.0858203$\times 10^{-7}$ & 4.8911238$\times 10^{-8}$\\
\enddata
\tablenotetext{\dag}{The Schott formula for the index of refraction is 
   $ n^2 = a_0 + a_1\lambda^2 + a_2\lambda^{-2} + a_3\lambda^{-4} + a_4\lambda^{-6} + a_5\lambda^{-8}$, 
where $n$ is the index of refraction and $\lambda$ is the wavelength in micrometers.  The coefficients are determined for operating conditions of 2~$^{\circ}$C and 456 mmHg}
\end{deluxetable*}

The camera lenses were fabricated in the optical shop of the UC Lick Observatory by David Hilyard and AR coated at ECI (Willow Grove, PA).  Each is mounted in a machined metal cell and secured by a continuous bead of room temperature vulcanizing silicone (Dow Corning RTV560, \citet{fata98}) designed to passively compensate and minimize stress on the lenses due to the temperature difference between the laboratory assembly conditions in Santa Cruz ($\sim 22$~$^{\circ}$C, 760 mmHg) and the operating conditions on Mauna Kea ($\sim 2$~$^{\circ}$C, 456 mmHg).  A machined spacer between elements 7 and 8 is slotted to accommodate the camera shutter (the second of the two identical Bonn shutters in the instrument).

The principle of the mount design, similar to that used for the DEIMOS camera (\citet{mast98}), requires the coefficient of thermal expansion (CTE) of the lens to be smaller than the CTE of its metal cell, and the CTE of the RTV to be larger than both the lens and cell material. When the camera is brought from lab temperature to operating temperature, the cell contracts more than the lens. The thickness of the RTV bond used to hold the lens into the cell is determined such that the thermal contraction of the RTV is the same as the difference between the contraction of the lens and cell. The cells are made from either 303 stainless steel (most of the glasses) or 7075 aluminum (CaF$_2$, S-FPL51Y) in cases for which the optical material has a large CTE.

Lenses were positioned in each cell on a rotating table with a series of gauges, and when the lens was centered adequately (the requirement is 15~microns for the finished cell), the RTV was injected into the gap and allowed to cure.  The lenses were mounted into front and rear stacks, with each cell shimmed into position to meet the alignment requirement.  Element 5 was delivered last from the coating vendor and completed the assembly.

\begin{figure*}
\plotone{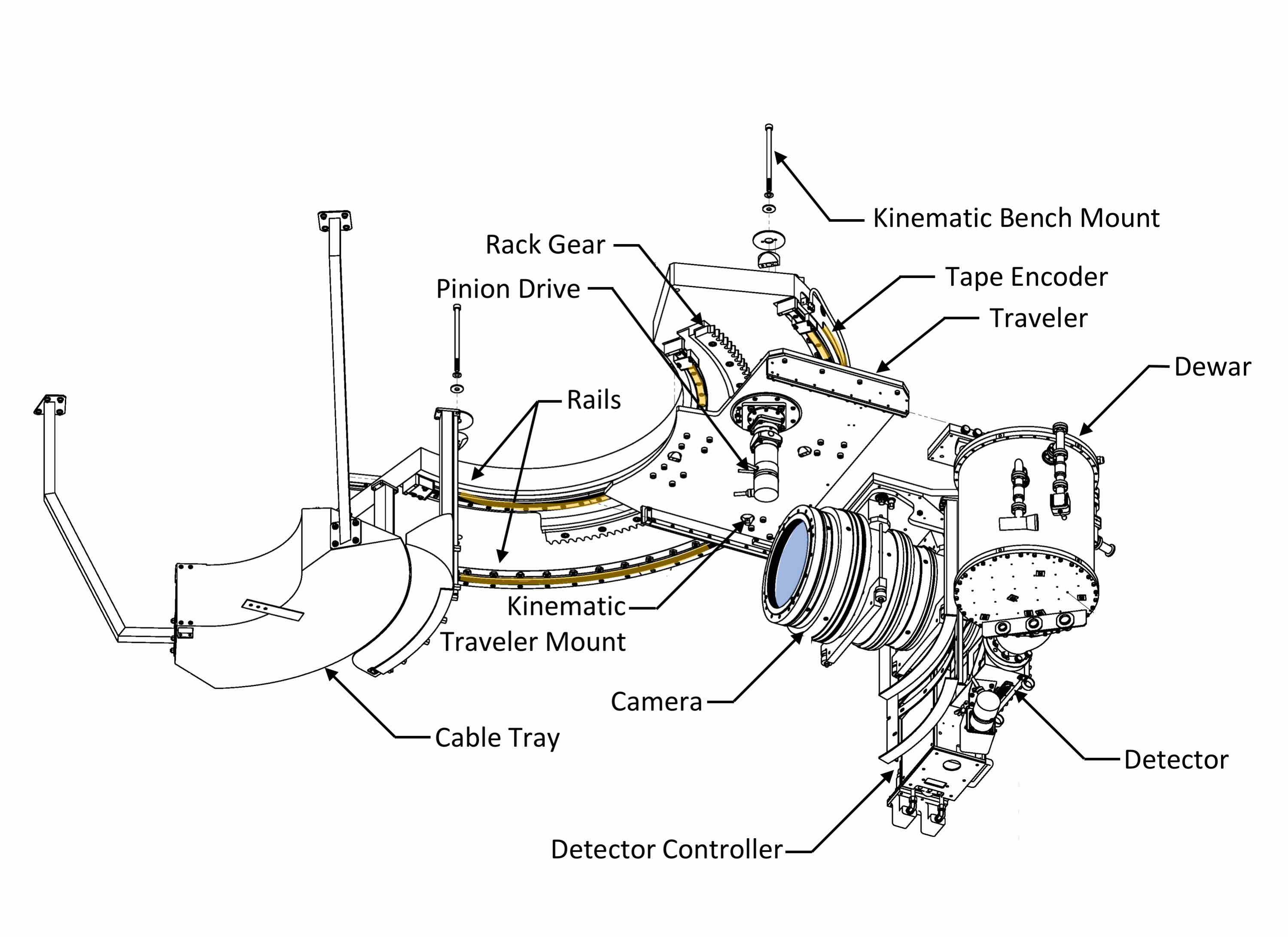}
\caption{The KCWI articulation mechanism assembly, which includes the camera and detector.  The stage is made from steel and is kinematically mounted to the underside of the optical bench.  A rack and pinion drive with magnetic tape encoder enables wavelength selection as well as access to a liquid nitrogen fill position.  Station keeping is provided by a servo motor mounted to the traveler stage.\label{f:art}}
\end{figure*}

The camera was delivered to Caltech where it was installed onto a steel articulation stage that carries both the camera and the detector.  A diagram of the assembly is shown in Figure~\ref{f:art}.  The traveler is articulated on curved rails that are centered around the grating axis of rotation and secured by a large custom stage that is kinematically mounted under the optical bench.  Wavelength selection is accomplished by means of a rack and pinion mechanism.  A servo motor mounted to the traveler provides station keeping during observations by means of a long magnetic strip encoder mounted to the perimeter of the stage.

\subsection{Detector}

\begin{figure}
\plotone{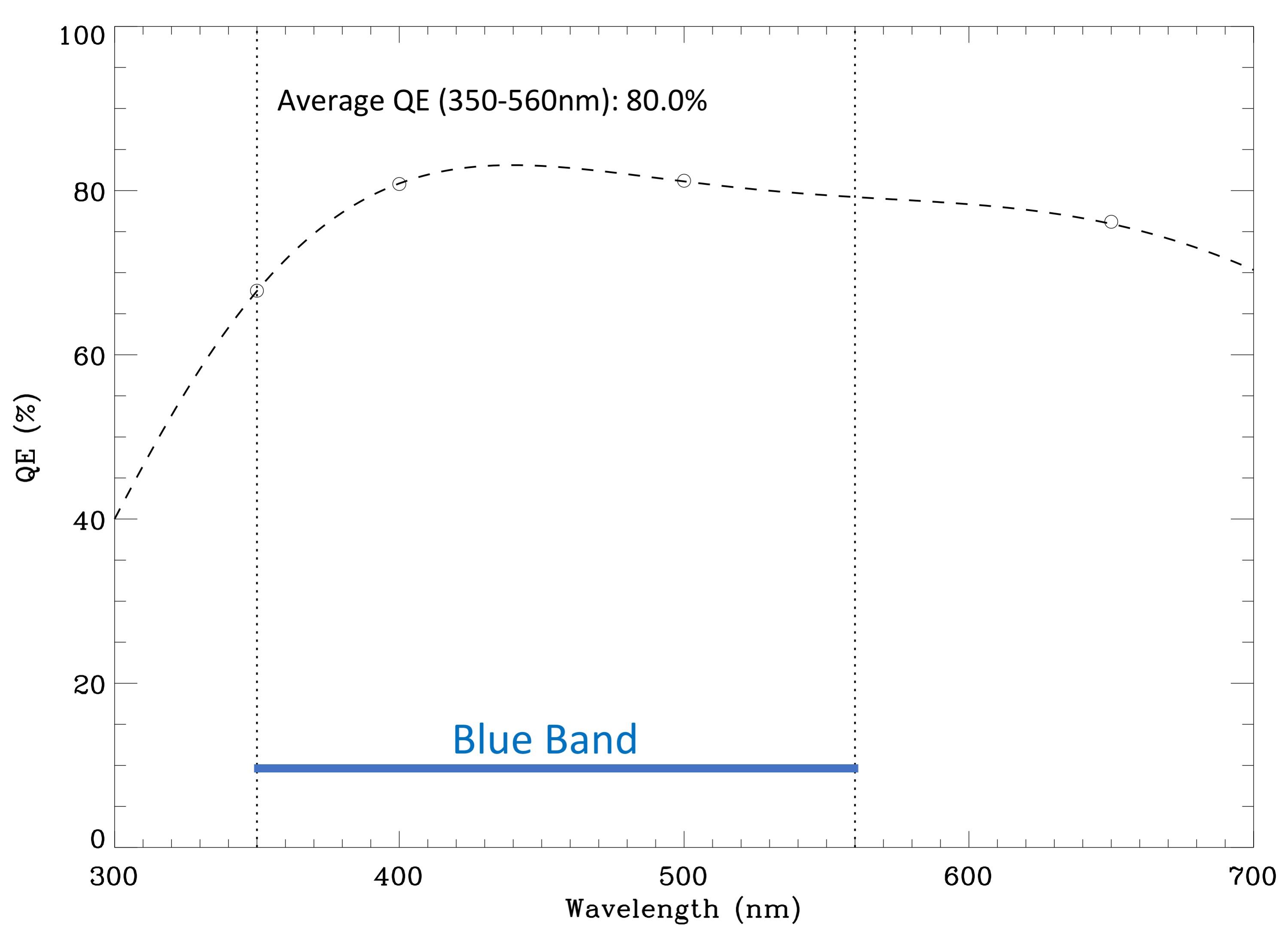}
\caption{The QE of the KCWI CCD is shown for four points measured by the vendor.  The dashed line is a fit to a standard Teledyne-e2v QE curve scaled to the measured values.\label{f:qe}}
\end{figure}

The camera is coupled to a liquid nitrogen cooled, AR-coated,
back-illuminated CCD detector. The last lens of the camera serves as
the vacuum window. The CCD is operated
between $-100 ^{\circ}$C (lab) and $-110 ^{\circ}$C (observatory), with a measured hold time of 96
hours at the summit.  The operating temperature is maintained by a
Lakeshore 336 controller to within $\sim0.1 ^{\circ}$C RMS. The dewar is filled from a location at the end of the articulation stage travel that engages a fill port and vent with externally accessible plumbing.  The vacuum is maintained by a charcoal getter located under the dewar and monitored by a 2~l-s$^{-1}$ ion pump.  Typical operating pressure when the dewar is cold is $<10^{-6}$~Torr, enabling the system to operate continuously for up to one year.  A focus mechanism provides precise ($\pm 1$~micron) piston control of the detector vacuum enclosure behind the camera barrel
through a 6~mm design range with the liquid nitrogen dewar held fixed and connected through a flexible bellows.  The CCD is a
Teledyne-e2v CCD231-84 that has $4096 \times 4112$ pixels with $15~\mu$m
pitch mounted on a SiC package with a pair of permanently attached and thermally isolating flexible ribbon cables. The surface passivation and coating is the blue-optimized ``astro broadband'' process, which provides enhanced QE below 400~nm.  The QE of the detector was measured at Teledyne-e2v and is shown in Figure~\ref{f:qe}.  The average value is 80\%.

\begin{figure*}
\plotone{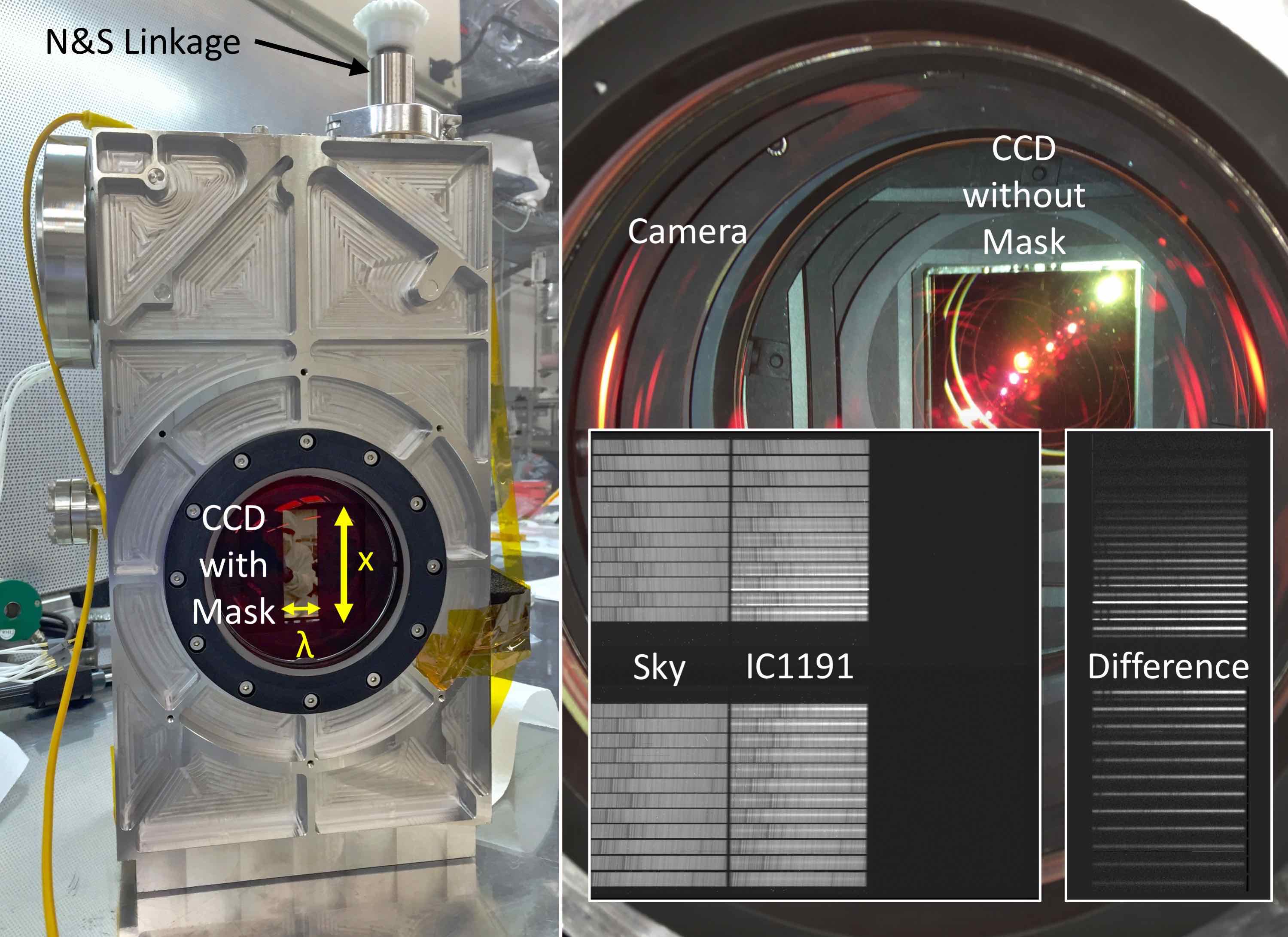}
\caption{An illustration of the deployable N\&S mask.  In the benchtop image at left, the mask is shown in the N\&S position over the CCD.  This is the mode used for the demonstration observation of IC1191 from 16-17 May 2017 shown in the inset on the right.  The night was moonlit and the image shows considerable sky background filling the slicer in addition to the spectra of the galaxy.  In the background on the right, the CCD is shown through the camera lens with N\&S mask in the OPEN position.  Note the red reflections in the lens that are due to the high out-of-band reflectivity of the camera coatings.  The red blocking filter eliminates this out-of-band light during ordinary circumstances.\label{f:ccdns}}
\end{figure*}

A deployable nod and shuffle (N\&S) mask inside the vacuum provides enhanced observational capability to KCWI.  The mask is made from stainless steel and is finished with black chrome.  It is mounted approximately 1~mm above the CCD on a pair of linear bearings and is actuated by an external ground-isolated stepper motor linked to a lead screw through a ferro-fluidic vacuum feedthrough.  The mask has three positions: open, dark and N\&S; its operation is illustrated in Figure~\ref{f:ccdns}.  On the left side of the figure, the detector housing is shown with mask in the N\&S position.  In this mode, the middle third of the CCD is exposed, truncating 2/3 of the bandwidth but enabling nearly simultaneous target and sky measurements.  During a typical observation, the telescope is ``nodded'' back and forth between a target and sky position for 2~minutes each.  When the instrument is finished with one target, the shutter is closed and the image is ``shuffled'' under the mask.  The telescope then points at the second field and repeats the process, typically for a total observation length of 1~hour (1/2 sky and 1/2 target).  An example of a N\&S spectrum of the galaxy IC1191 is shown in the inset on the right side of the Figure.  The observation of the sky and the galaxy are both on a single CCD image (with only one readout noise penalty).  This 20~minute demonstration was taken on the moonlit night of 16-17 May 2017 during KCWI commissioning with the large slicer and BL grating centered at 450~nm.  While the N\&S capability enables superior sky subtraction for extended objects, in cases where the sky can be adequately measured from within the field of view, the bandwidth can be extended by moving the mask to the OPEN position.  

The detector controller is a standard Astronomical Research Cameras (San Diego, CA) Generation III controller designed by Robert Leach with quad channel, 16 bit readout and correlated double sampling.  The read noise is $\lesssim 3$ electrons with a controller gain of $0.16$~electrons-DN$^{-1}$ ($\sim 4.5$~electrons in fast mode) and essentially unchanged from the measured value in the laboratory at Caltech.  We attribute this to our attention to ground isolation in the detector system.  Care was taken throughout the design process to eliminate the possibility of ground loops by ensuring the CCD (including the package) is only grounded to the detector controller.  We have measured the equilibrated dark current to be 1~electron-pixel$^{-1}$-hour$^{-1}$ at -110~$^{\circ}$C. Care should be taken that the dark current is elevated (a known charge trapping effect) for several hours after a controller restart ($\sim 4$~electron-pixel$^{-1}$-hour$^{-1}$ after 2 hours, $\sim 2$~electron-pixel$^{-1}$-hour$^{-1}$ after 6 hours).

The controller is mounted next to the CCD vacuum housing in order to minimize the cable length from the CCD to the video input (estimated to be approximately 30~cm including the flex-cable attached to the CCD itself).  The controller is connected by optical fibers to a Linux host computer installed in the environmentally controlled data room at the observatory summit.  The controller software supports readout from any of the four amplifiers, or from either serial register.  Two readout speeds are also available, as is 2x2 binning.  In order to optimize read noise performance, high gain is used that limits the full well to the range of the A/D converter.  Readout times for the available modes range from 7~s ($2\times 2$ fast, quad amplifier) to 337~s ($1\times 1$ slow, single amplifier) with a typical science configuration ($2\times 2$ slow, single amplifier) reading out in 106~s.



\section{Coatings}
\label{s:coatings}

The KCWI design contains a meniscus window, nine mirrors, two filters, a grating and a nine element camera lens.  Careful attention to the design, performance and stability of the coatings is clearly essential to achieve the instrument requirements.  Throughout the design process, we made every effort to ensure a high performance red channel could be added without recoating existing elements.  Table~\ref{tbl:coatings} shows the coatings and design bands used in KCWI, by element and by vendor.

\begin{deluxetable*}{llcl}
\tablecaption{KCWI Coating Overview\label{tbl:coatings}}
\tablehead{
\colhead{Element} & \colhead{Coating} & \colhead{Design Band} & \colhead{Vendor}\\
\colhead{} & \colhead{} & \colhead{(nm)} & \colhead{}
}
\startdata
Window            & Dielectric    & 350-1050 & ECI (Willow Grove, PA)\\
K-mirror          & Dielectric    & 350-1050 & ECI\\
Guider Dichroic   & Dielectric    & 350-630\tablenotemark{a}  & Asahi (Japan)\\
IFU               & Multilayer-Ag & 350-1050 & Optics Balzers (Liechtenstein)\\
Collimator        & Multilayer-Ag & 350-1050 & Zecoat (Torrance, CA)\\
FM1               & Multilayer-Ag & 350-1050 & Zecoat\\
FMD               & Enhanced-Al   & 350-560  & Clausing (Skokie, IL)\\
FM3               & Enhanced-Al   & 350-560  & Clausing\\
Blocking Filter   & Dielectric    & 350-560  & Asahi\\
Camera            & Dielectric    & 350-560  & ECI\\
\enddata
\tablenotetext{a}{The guider dichroic is designed to permit HeNe laser light at 632.8~nm into the blue channel for alignment purposes.}
\end{deluxetable*}

\subsection{Reflective Coatings}

The common optics of the eventual red and blue channels include the window, three K-mirror reflections, two IFU reflections, the collimator and the cylindrical FM1 mirror (the guider will be redesigned when the red channel is added).  The existing FMD fold mirror will be replaced by a dichroic, and all existing elements in the blue arm from the FM3 mirror onward will not see light longer than 560 nm.  We invested in broad-band dielectric and multilayer silver coatings to boost red throughput for the common reflective elements, while the blue arm mirrors are coated with enhanced aluminum or multilayer dielectric.  

The three elements of the K-mirrors represent a necessary addition to the mainframe optical design in order to accommodate the Nasmyth port location.  Since this mechanism increases the number of required instrument reflections by 50\%, it is important to minimize its impact on performance as much as possible.  Furthermore, the alignment of the K-mirror is a precise and challenging operation that incentivizes the inclusion of robust coatings.  As such, we elected to invest in broad-band dielectrics from ECI, and each provides average reflectivity $\gtrsim 99$\% over the 350-1050~nm band.  The K-mirror substrates are Zerodur with a diameter-to-thickness ratio of approximately 10.  The mirrors were each back-side polished and stress-relieved with a compensating coating on the back to preserve the front side surface figure.  We found that the finished mirrors exhibited a mainly spherical $0.5$~wave peak-to-valley surface error due to residual coating stress as measured by an interferometer, well within the $3$~wave requirement and not severely degraded from the initial $0.1 - 0.2$~wave fabrication quality.

The reflective dielectrics provide similarly high {\em s-} and {\em p-} polarization reflectivity, however the complex interaction of light with the rotating K-mirror causes a phase shift of the incoming beam.  We investigated this property by illuminating the instrument with linearly-polarized light from a flat field source in the laboratory.  For each K-Mirror angle, we analyzed the transmitted light using a selectable downstream polarizer that is part of the calibration unit.  We find that as the K-Mirror rotates, the transmitted beam polarization changes character from linear to circular and back to linear again, where the transmitted beam is linearly polarized at K-Mirror rotation angles of $0^{\circ}$ and $90^{\circ}$, but circularly polarized in between at $45^{\circ}$.  Since this is a slowly varying effect, the N\&S mode can be employed for demanding applications. Performance simulations with variable beam polarization were included in \citet{morrissey12}.

The remaining blue and red common-path optics, the COL and FM1, are coated with multi-layer-protected silver for enhanced broad band performance.  A significant effort was expended in optimizing these specialty coatings for the KCWI application taking advantage of the experience of the team at Lick Observatory and in industry (\citet{phillips12,sheikh08}).  Non-tarnishing silver coating designs have been developed over the past fifty years with success. Denton Vacuum's FS-99, protected by Al$_2$O$_3$/SiO$_2$, and Lawrence Livermore Laboratory's nitride-based protective designs are noteworthy examples (\citet{adams72,wolf00}), and these coatings have been applied to many ground and space-based mirrors. Recently, the desire to develop a silver reflector for the UV has led to new research (\citet{sheikh16}). The silver coatings are really dual designs in the sense that the multilayer protective coating is providing most of the reflectivity short of 400~nm, while the silver itself is an excellent coating for longer wavelengths (and far superior to aluminum in the I band).  Nonetheless, unprotected silver is well known to be at risk of tarnishing on exposure to the elements. Since lifetime is an important consideration both to manage overall instrument cost and to accommodate the future red channel, we embarked on a rigorous test program for the KCWI silver coatings that included wiping with water and solvents as well as a 10 day accelerated lifetime test in a 80~$^{\circ}$C/80\%~RH salt atmosphere (\citet{phillips16}).  A variety of designs based on protective metal-oxides, metal-nitrides, and metal-fluorides were tested and
compared.  We found that while no designs survived our abusive test without any damage at all, some fared substantially better than others.  Of particular note we found that a water drop (although not a solvent wipe) deposited on any coated mirror surface prior to the lifetime test would always reveal itself in the final results as a damaged area of the coating, pointing to the importance of a dry instrument purge at the observatory.  The final design ZeCoat samples were among the best of all of those that were evaluated, and revealed only minor degradation after environmental testing.  In the end we chose two proprietary designs for the protected silver in KCWI: the IFU was coated in Europe by Optics Balzers AG (Liechtenstein), while the two large mirrors were coated at ZeCoat Corporation (Torrance, CA).  The average reflectivity of both designs is close to 95\% in the blue band.  The ZeCoat design received the most optimization and performed the best in the lifetime test with salt atmosphere. The Optic Balzers coating passed a similar heat/high humidity test but without the salt.  The project decided that the cost, export requirements and transportation risk involved in coating the IFU in the U.S.A. outweighed the performance margin of the optimized coating, and we therefore accepted some increased risk on the IFU coating lifetime, judged to be reasonable in view of the fact that samples with similar Balzers coatings were remeasured with no degradation after five years in ``desk drawer'' conditions at Winlight.

\begin{figure*}
\plotone{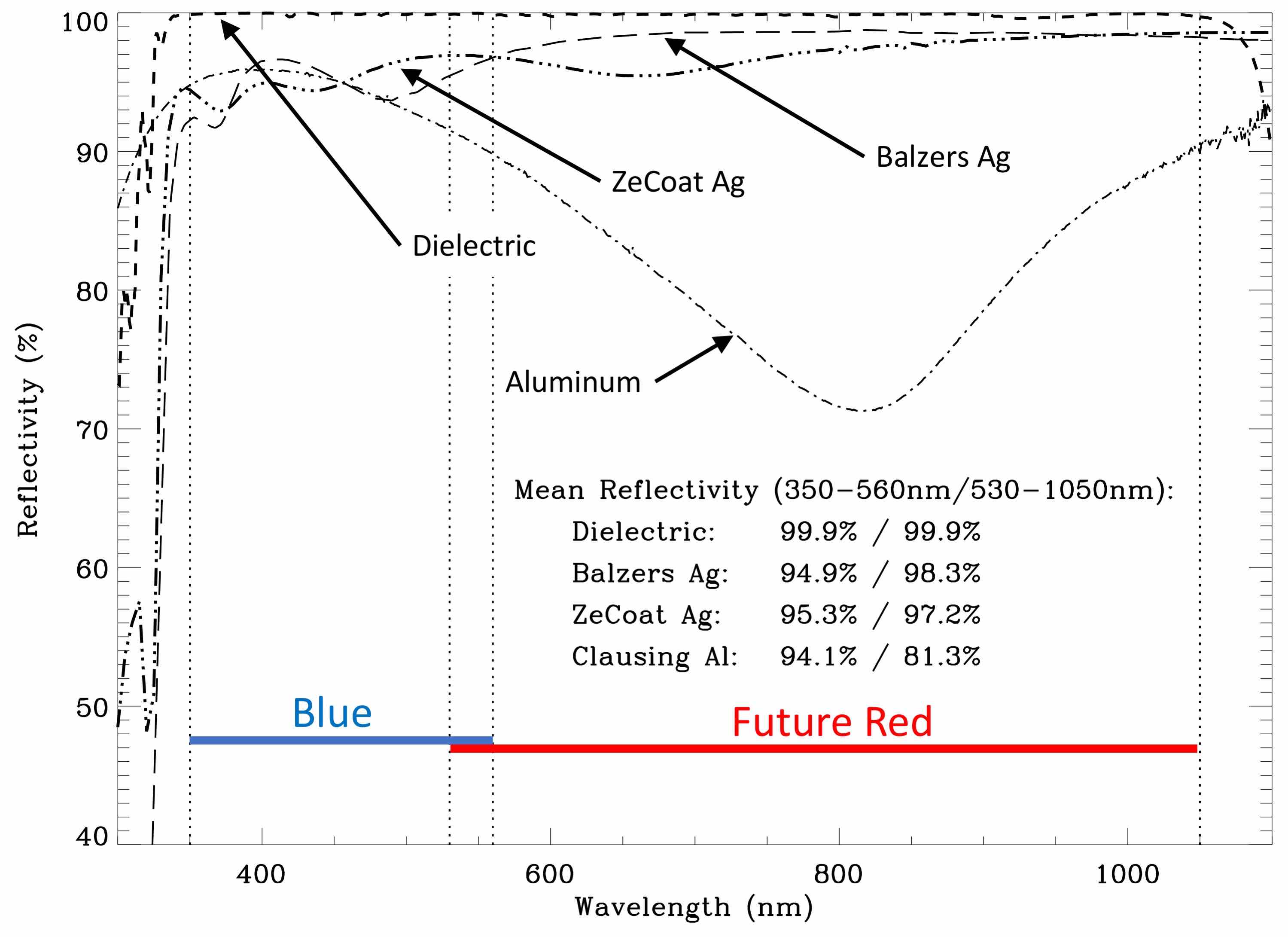}
\caption{The measured reflectivity of the KCWI mirrors illustrating both current blue band performance and the potential for high performance in the future red arm.\label{f:lrgoptref}}
\end{figure*}

Downstream of the FM1 cylinder, the FMD/FM3 pair of relay mirrors are coated with a proprietary enhanced-aluminum (``BB-100'') coating from Clausing (Skokie, IL) that has been optimized for the 350-560~nm band.  The average reflectivity of these mirrors is 94\% in the blue band.  The measured results for all of the KCWI reflective optics are shown in Figure~\ref{f:lrgoptref}.  It is apparent that the dielectric coatings are best of all, while aluminum is an excellent cost-effective option in the blue and silver can effectively be used to coat large surfaces with enhanced red performance.

\subsection{Windows and Filters}

\begin{figure}
\plotone{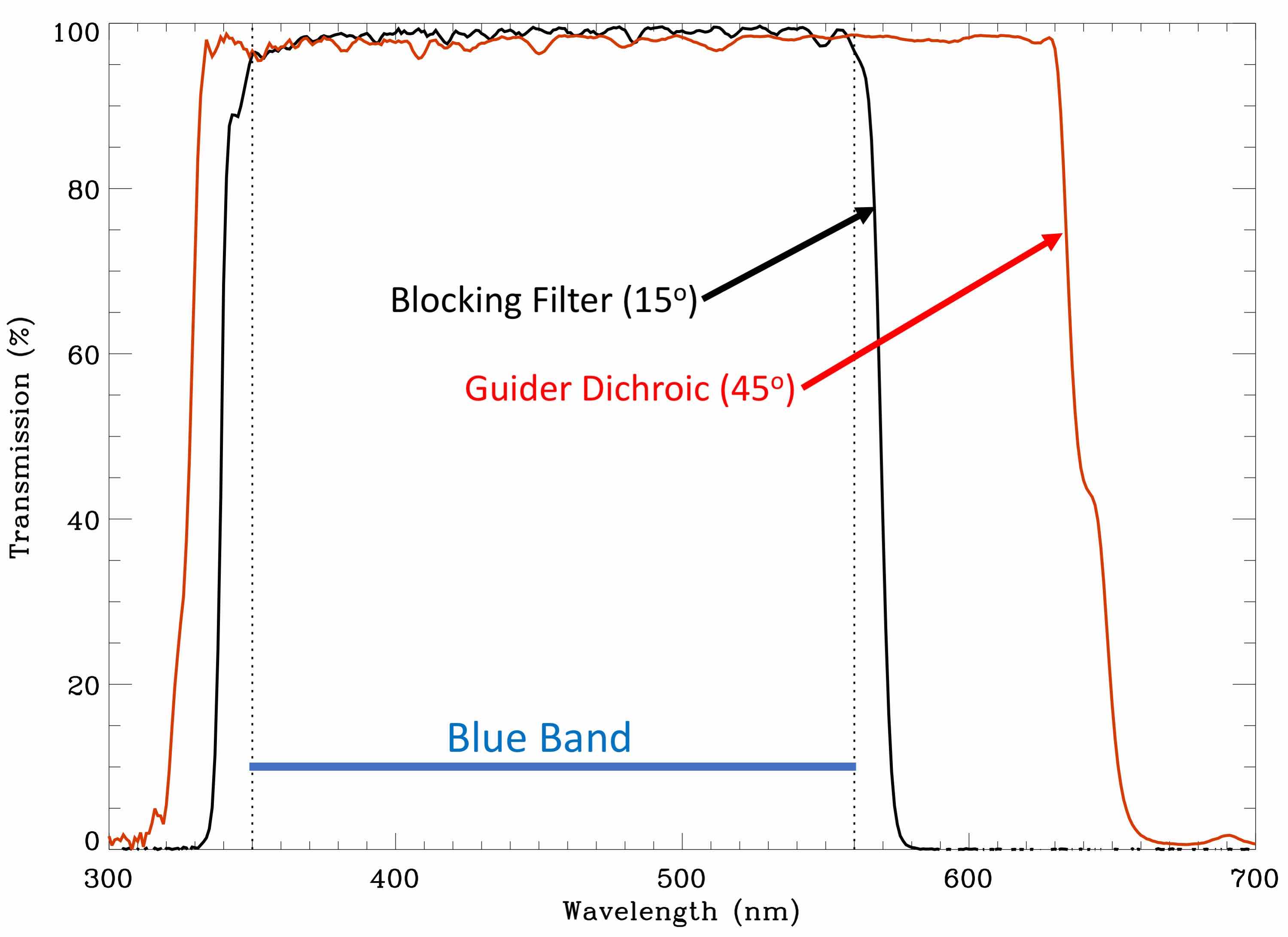}
\caption{{\em Black line:} The measured throughput of the red blocking filter. {\em Red line:} The measured throughput of the guider dichroic, which is designed to permit HeNe laser alignment tools.  The overall instrument band can be broadened by removing the blocking filter at the expense of greatly increased ghosting in the camera.\label{f:dichroref}}
\end{figure}

A fused silica meniscus window protects the entrance of the spectrograph in front of the K-mirror.  It is coated with a standard commercial (ECI) broad-band AR coating averaging 1.5\% reflectivity per surface.  Downstream of the K-mirror, the guider dichroic transmits 97\% of the 350-630~nm blue band light while reflecting 98.5\% of the 650-900~nm red light to the MAGIQ guider.  The intermediate choice of cutoff wavelength in the guider dichroic permits the use of HeNe laser alignment tools inside the spectrograph.  A camera red blocking filter ensures only blue-band light enters the grating and camera.  It is coated with a dielectric filter from Asahi (Japan) and tilted $15^{\circ}$ to eliminate the possibility of ghost light from the camera return beam.  It should be noted that the spectrograph can be used with wavelengths as red as the guider dichroic 630~nm cutoff by removing the red blocking filter, although ghosts in the camera (discussed in the next section) would be expected to be relatively bright in this configuration.  The measured throughput of the guider dichroic and camera red blocking filter are shown in Figure~\ref{f:dichroref}.

\subsection{Camera Coatings}

The last element in the spectrograph, the nine element camera, also has a major potential impact on the overall instrument performance.  Eighteen surfaces plus the detector have the opportunity to scatter light and to re-direct it in undesirable ways.  We used the Zemax (Kirkland, WA) ghost focus generator to produce a Gaussian estimate of the expected ghosts resulting from double bounces through the camera system.  The CCD is included as an element, and it produces some of the brightest ghosts because it is very reflective compared to any of the camera lens surfaces.  For modeling purposes, we used 30\% CCD reflectivity (worst case) and 0.5\% lens reflectivity to make conservative estimates.  There are two primary forms of ghost: reflections that move with the source (forming a halo), and others that move in the opposite direction as the image is scanned across the field.  Of the three brightest ghosts from the simulation, two involve the CCD and its nearest element, the field flattener, while the third is from a double bounce inside the field flattener (suppression due to the low coating reflectivity in this case is compensated by the close proximity to the CCD).  These three bright ghosts are between 2 and 20~mm diameter (occupying 3-30\% of the detector field of view) with {\em integrated} intensities as bright as 0.1\% of the source.  Large extended ghosts are exacerbated by dispersion, which integrates them while simultaneously attenuating the original point source.  The ghosts are quite faint, but must be evaluated relative to the sky continuum.  For example, simulations indicate that a magnitude 20 star forms a broad, diffuse composite ghost at around the $10^{-3}$ of the sky level, depending on the dispersion.  Observers should take care to avoid bright sources in regions for which the sky subtraction is required to work at the finest level (e.g. ``light bucket mode,'' for which a single spectrum is created from the entire slicer area).

In view of these results, we were careful to evaluate the performance of each coated element before integrating it into the camera.  At the low reflectivity levels we were evaluating ($\sim 0.2$\%), we found that humidity and age could degrade performance (from a combination of coating porosity and deterioration of the black coating deposited on the rear surface of each witness sample).  Adhesion was a challenge on some of the surfaces, particularly those made of CaF$_2$. Several required multiple attempts (with associated re-polishing) in order to meet requirements.  ECI chose two processes to deposit the AR coatings: electron beam and ion beam sputtering (IBS).  We found that the lowest reflectivity (due to the wider range of allowed materials) was achieved with the e-beam, but the IBS process yielded the best adhesion and repeatability, and was chosen for the most problematic elements.

\begin{figure}
\plotone{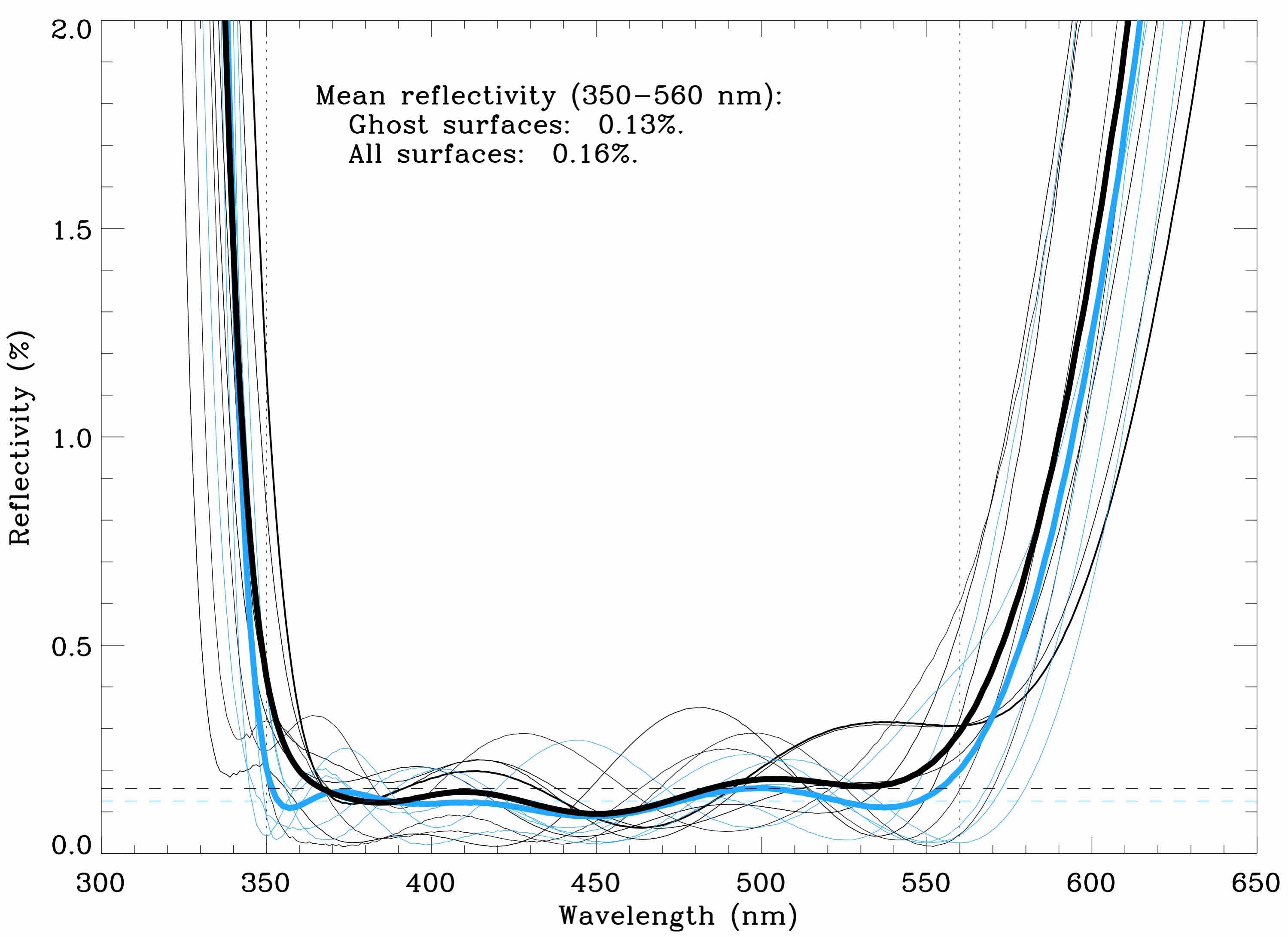}
\caption{The measured reflectivity of the eighteen camera lens surfaces.  Surfaces with the potential to generate bright ghosts are highlighted in blue.  The average reflectivity of all of the ``ghost surfaces'' is plotted in bold blue while the average reflectivity of all surfaces combined is plotted in bold black.  The mean reflectivity is under 0.2\%.\label{f:camr}}
\end{figure}

\begin{figure}
\plotone{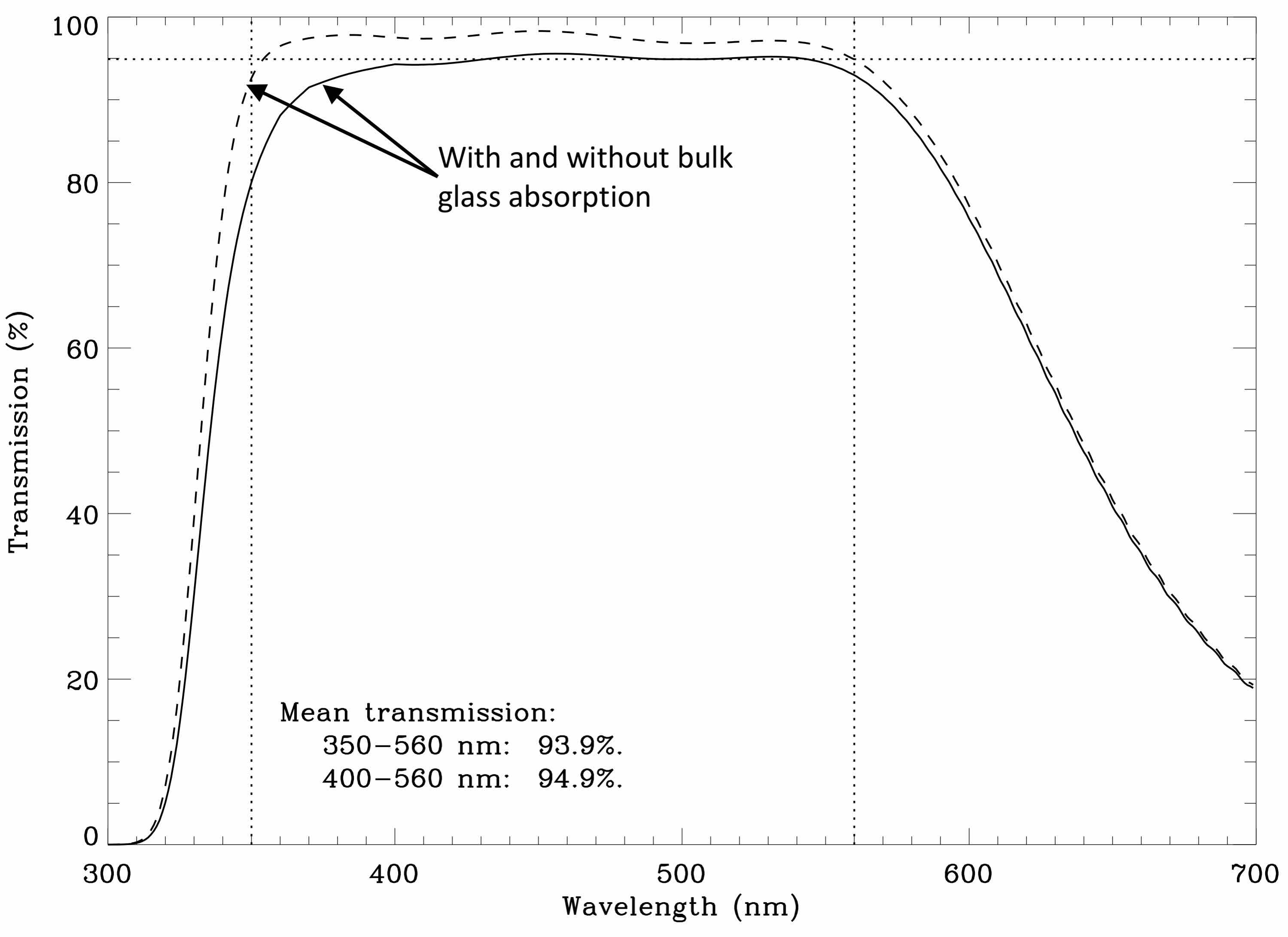}
\caption{{\em Solid line:} The model transmission of the KCWI nine element camera lens based on the measured reflectivity of all of the individual lens surfaces multiplied by a factor to account for bulk absorption in the glass.  {\em Dashed line:} The estimated camera lens transmission in the absence of bulk glass absorption.\label{f:camt}}
\end{figure}

The measured reflectivity results from our camera coating program are shown in Figures~\ref{f:camr}.  The results for the worst ghost generating surfaces are depicted in blue, revealing that on average we were able to hold their reflectivity to under 0.15\%, while the overall camera surface reflectivity is under 0.2\%.  When combined, these result (with the addition of some predicted glass absorption below 400~nm) in an overall camera transmission of over 96~\% as shown in Figure~\ref{f:camt}.

\subsection{Baffling and Stray Light}

The control of ghosts and stray light is an important element of the design of a background-limited instrument such as KCWI.  In the case of a reflective image slicer, a certain amount of shadowing and stray light is unavoidable due to the close proximity of sharp reflective edges.  Light incident close to the edge of one slice may collide with the wall formed by the adjacent slice and be mis-directed.  The KCWI design includes a range of slice angles spanning $\pm 27^{\circ}$ with a $\pm 4^{\circ}$ gap between the central two slices to allow room for light to enter from the telescope through the pupil array.  While the two central slices have the largest angular discontinuity ($8^{\circ}$ vs a more typical $2^{\circ}$), they have the minimum shadowing because the central elements of the pupil array are designed along a diagonal (see Figure~\ref{f:vslit}) and each accepts light from one of the two central slices in such a way that incident light is reflected away from the wall at the center.

\begin{figure}
\plotone{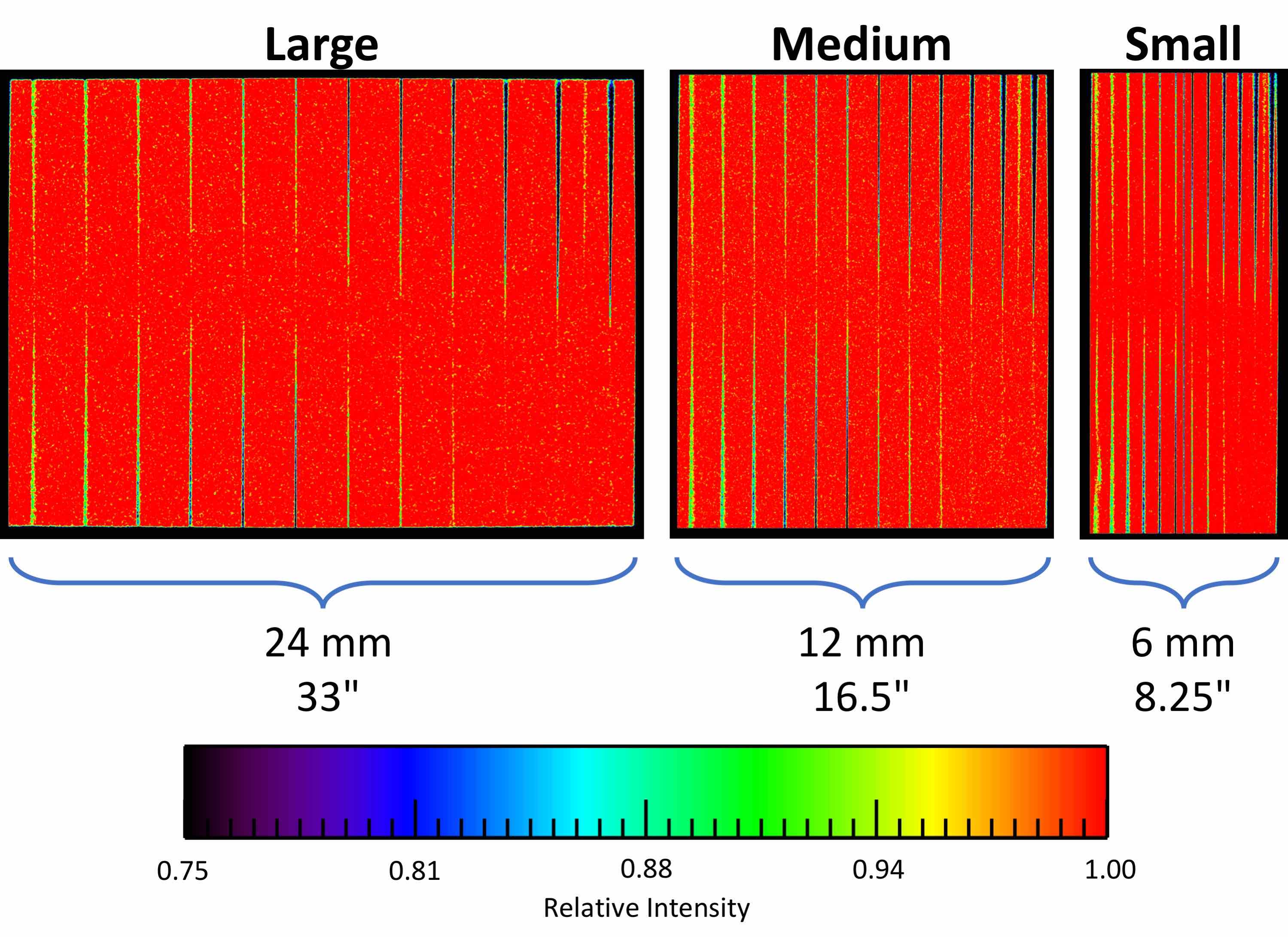}
\caption{Reconstructed flat field images of each slicer accounting for vignetting as determined with a non-sequential raytrace simulation.  Vignetting is correlated with slice edges as expected and ranges from a maximum of 5\% for the large slicer up to 25\% for the small slicer in isolated locations.  The effect is reduced when considered with a convolved point source, typically $>350 \mu$m FWHM.\label{f:slicerv}}
\end{figure}

We have performed a non-sequential analysis of the IFU with each image slicer design uniformly illuminated by a beam of light with a range of angles similar to the Keck telescope.  Rays are considered to have been vignetted if they are eliminated at any point in the instrument path, but the effect is dominated by internal reflections in each slicer.  We find that under these illumination conditions the large slicer vignettes a maximum of 5\% of the light from any location, while the medium and small slicers vignette a maximum of 9\% and 25\% respectively.  Results of this analysis are shown in Figure~\ref{f:slicerv} and show that the vignetting is concentrated at specific slice edges particularly toward the corners of the field.  The effect in the small slicer is mitigated by the fact that there are approximately two slices per seeing-limited PSF, and therefore is reduced by a factor of $\sim2$ when the footprint on the slicer is taken into account.

A fraction of scattered rays from each slice will land on a pupil mirror and be transmitted though the system.  The majority are eliminated by a series of baffles designed around the footprint of the virtual slit at each of the four large mirrors, the red blocking filter, and (most effectively) on both sides of each grating.  The remaining stray rays reach the detector and have the effect of broadening monochromatic spectral line images by approximately 1\% with 0.5\% of the total integrated flux.  Spectral features that are broader will reduce the magnitude of the effect proportionately.

During the design phase, we identified the contribution of a ``narcissistic ghost'' that resulted from light reflecting off of the detector and traveling backward through the optical system to re-image on the slicer. The return image is superimposed on the initial slicer
input, and the majority of this light is redirected back out to space. However, the return image is not of
the same quality as the input because it has traversed the spectrograph twice. The fuzzy return image
can therefore straddle adjacent IFU slices and be re-directed back into the spectrograph through the wrong channel.  This ghost (which now has a peak contribution of a few parts in $\sim10^{-3}$ in the central channels of the small slicer and much less in the others) was mitigated by adjusting the compound angle of the two central slices to move the two central pupil mirrors outboard approximately 1~cm as can be seen in Figure~\ref{f:vslit}.  The effect of this is to re-direct the return light from the system that mixes the two central channels off the IFU pupil array and out of the optical path.

\subsection{Predicted Optical Throughput}

\begin{figure}
\plotone{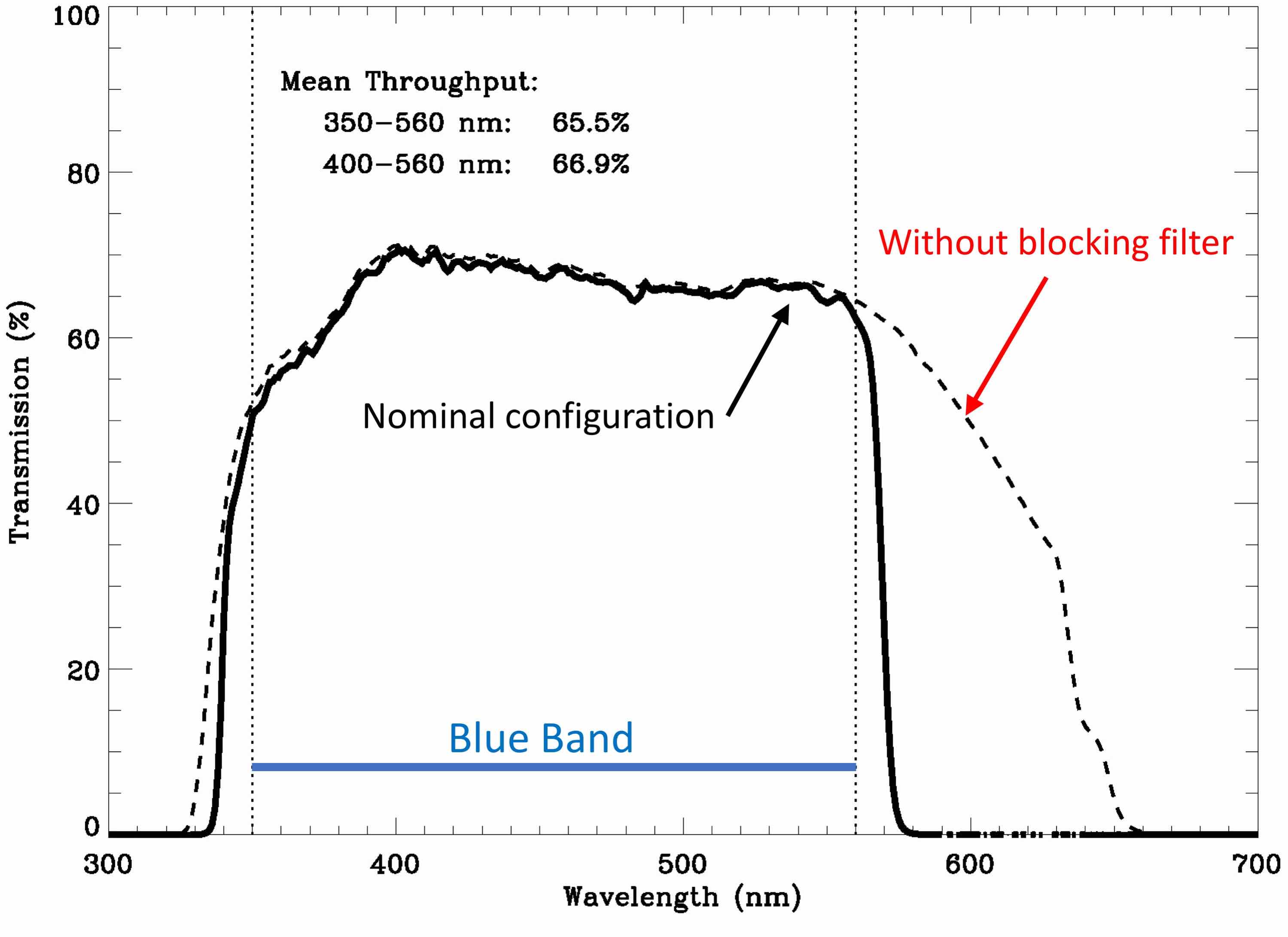}
\caption{{\em Black line:} The model optical throughput of KCWI based on the measured reflectivity and transmission of the guider dichroic, nine mirrors, red blocking filter and camera.  The detector and grating efficiencies are not included.  {\em Dashed line:} The model throughput with blocking filter removed (at a cost of increased camera ghosting).\label{f:instnograting}}
\end{figure}

The overall instrument throughput, which is the product of the reflectivities of the nine KCWI mirrors and the transmission of the window, dichroic and red blocking filter, is shown in Figure~\ref{f:instnograting}.  The rolloff shortward of 400~nm is primarily due to absorption in the camera glass.  The average throughput from 350-560~nm is 65.5\%.  This figure does not include the effect of the grating or detector in order to isolate the performance of the coatings.  We would anticipate similar high throughput in the planned 530-1050~nm red channel based on witness sample measurements.

\subsection{Predicted Optical Resolution}

\begin{figure}
\plotone{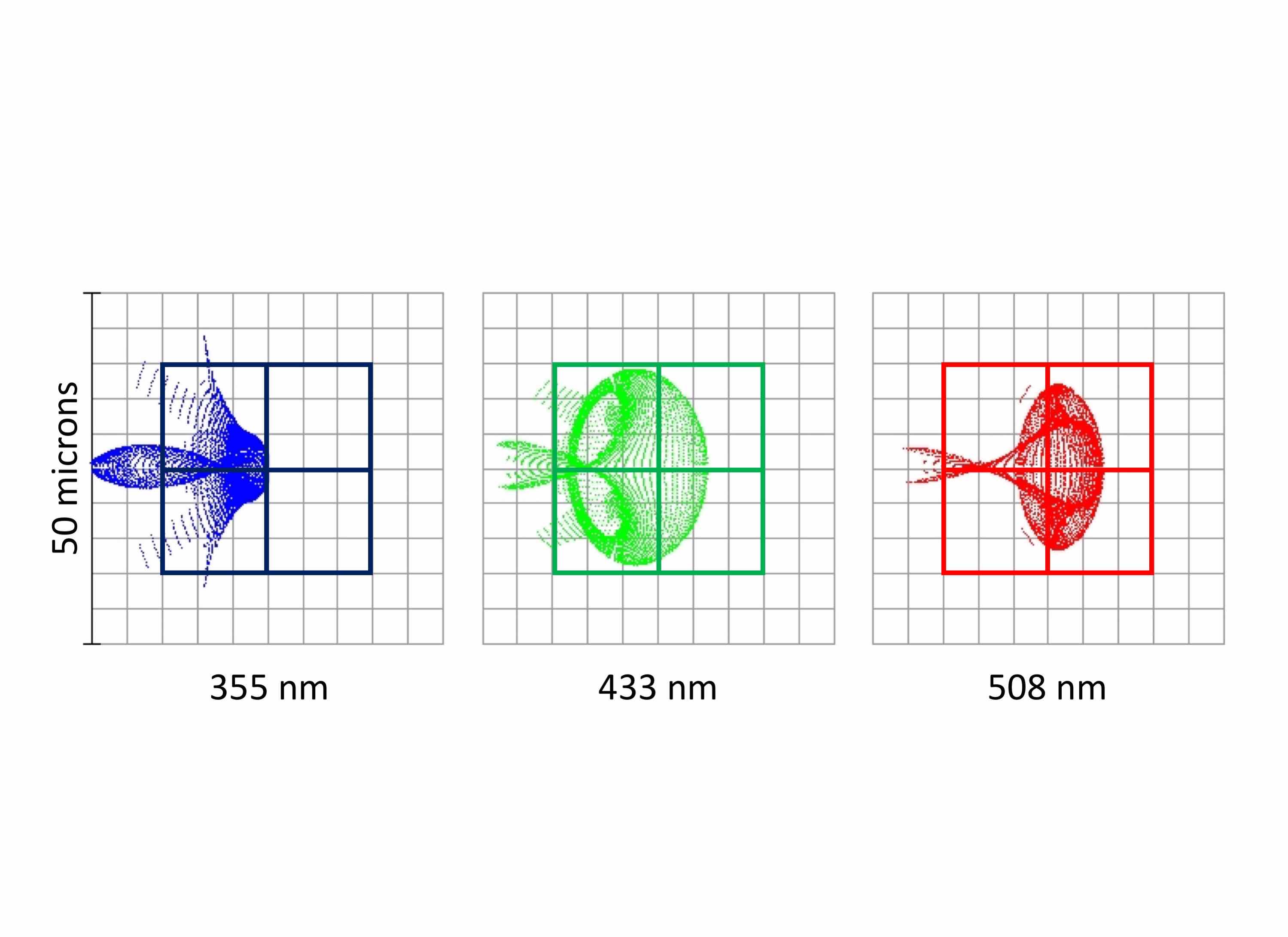}
\caption{Model end-to-end spatial resolution of KCWI with a BL grating at three wavelengths for a point source input.  The $2\times 2$ grid in the center of each image depicts the pixel sampling with $1\times 1$ binning.\label{f:spatialreszmx}}
\end{figure}

The KCWI design achieves pixel-sample-limited resolution without re-focus as shown in Figure~\ref{f:spatialreszmx}.  The cylindrical FM1 corrector ensures that the performance of the system is completely dominated by the large 305~mm focal length camera lens.  Good seeing at the summit of $0.5\arcsec$ corresponds to about $3.5$ pixels at the detector, while the highest resolution small slicer slit image is nearly Nyquist sampled at 2.3 pixels.  Optical imaging performance is comparable with all grating and filter combinations, but note that the small slicer oversamples the seeing-limited point spread function at the focal plane.

\section{Data Reduction Pipeline\label{drp}}

The data reduction pipeline (DRP) was developed using data from the PCWI, which is optically similar to the large slicer configuration of KCWI.
Here we will
provide the basic outline of the KCWI DRP; a more detailed description of the
algorithms used will be presented in \citet{neill18b}.

The DRP is written in the Interactive Data
Language (IDL)\footnote{Harris Geospatial Solutions; Herndon, VA}.  Versioning and distribution is handled through an on-line
repository (\citet{neill18a}).  Key features include:
\begin{enumerate}
\item Modular processing stages with minimum resampling for wavelength calibration, flat fielding, etc.
\item User-editable files for control of how the DRP operates
\item Detailed observation and configuration information stored in each image header.
\item Capability to process an entire night of data and to select sets of calibration images for each science observation
\end{enumerate}

\subsection{Pipeline Overview}

The pipeline is divided into nine modular stages as shown in Table~\ref{tab_drp_stages}.
Each of these stages 
produces a detailed log file and complementary ancillary diagnostic outputs.

Table~\ref{tab_drp_products} lists the primary image products for each
stage of the DRP.  The filenames are constructed as follows:
{\tt kbYYMMDD\_NNNNN\_sfx.fits},
where {\tt YYMMDD} is replaced by the UT date of observation,
{\tt NNNNN} is replaced by the zero-padded image sequence number,
and {\tt  sfx} is replaced by the file suffix listed in the
Table~\ref{tab_drp_products}.

\begin{figure*}
\plotone{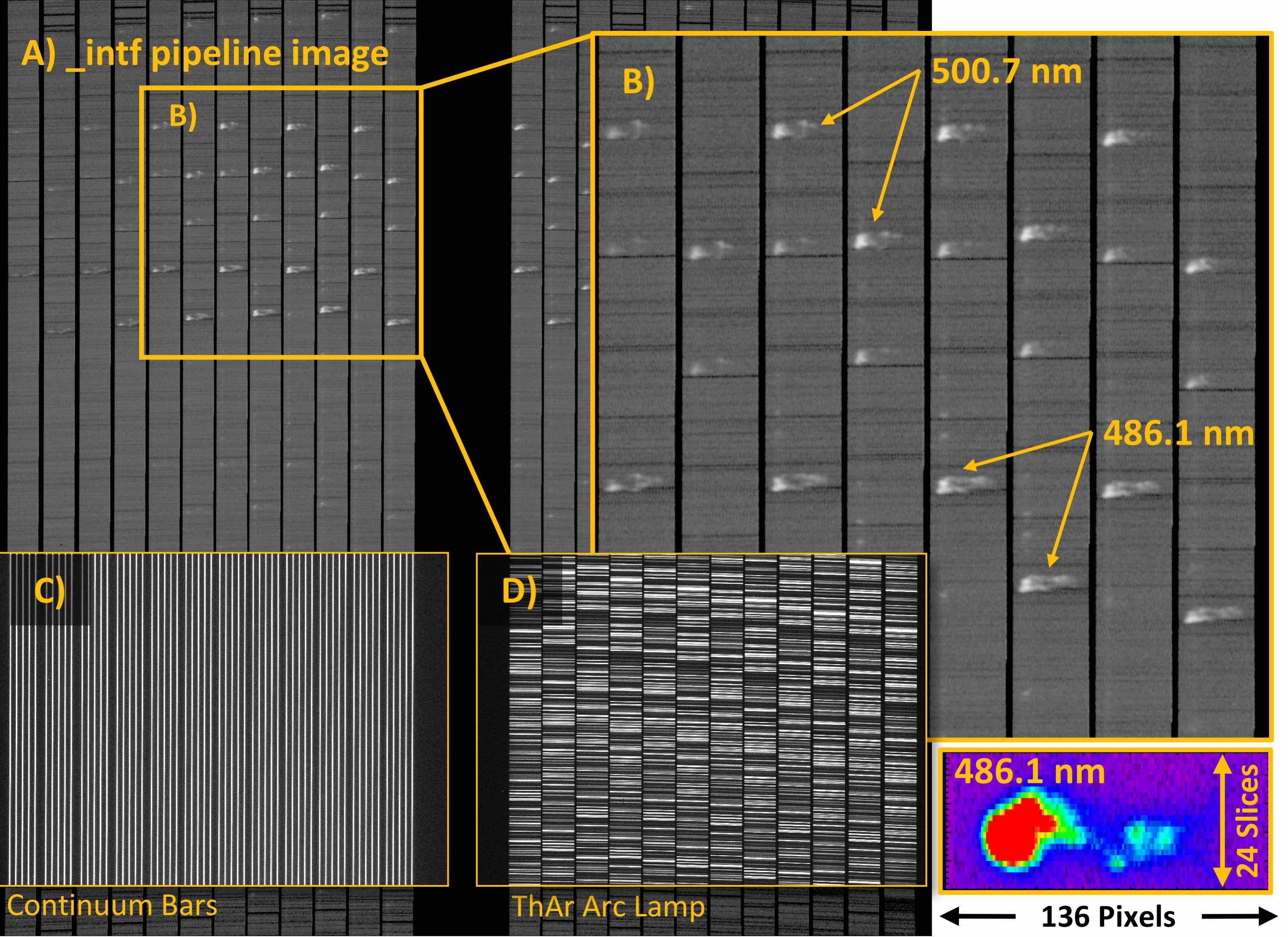}
\caption{A sample of pipeline image reconstruction.  {\it Panel A:} a log-stretch
flat fielded image ($\_$intf) of HH32 taken with the small slicer and BM grating centered at 470~nm.  {\it Panel B:} A magnified view of some emission lines revealing the distribution of different species.  {\it Panel C:} An LED-illuminated ``continuum bars'' calibration image. {\it Panel D:} A ThAr arc lamp flat field calibration image.  {\it Lower Right:} A pipeline-reconstructed image of the small slicer at a single wavelength.}
\label{fig:drp_images}
\end{figure*}

Figure~\ref{fig:drp_images} provides a composite of several of the science and calibration images typically associated with the data reduction. The Panel A shows a 10 minute observation of the star forming region HH32 taken during commissioning (small slicer with BM grating centered at 470~nm) that has been processed through Stage 4 (flat field) of the pipeline.  Panel B shows an expanded view of two prominent emission features.  The image is overlaid (Panels C and D) with insets showing a calibration continuum lamp bar image used for spatial alignment of each spectrum and a ThAr arc lamp image used for the wavelength alignment.  Fraunhofer lines are visible across the log stretch image.  The lower right panel shows a pipeline reconstruction of the image on the slicer at one wavelength.

\begin{figure*}
\plotone{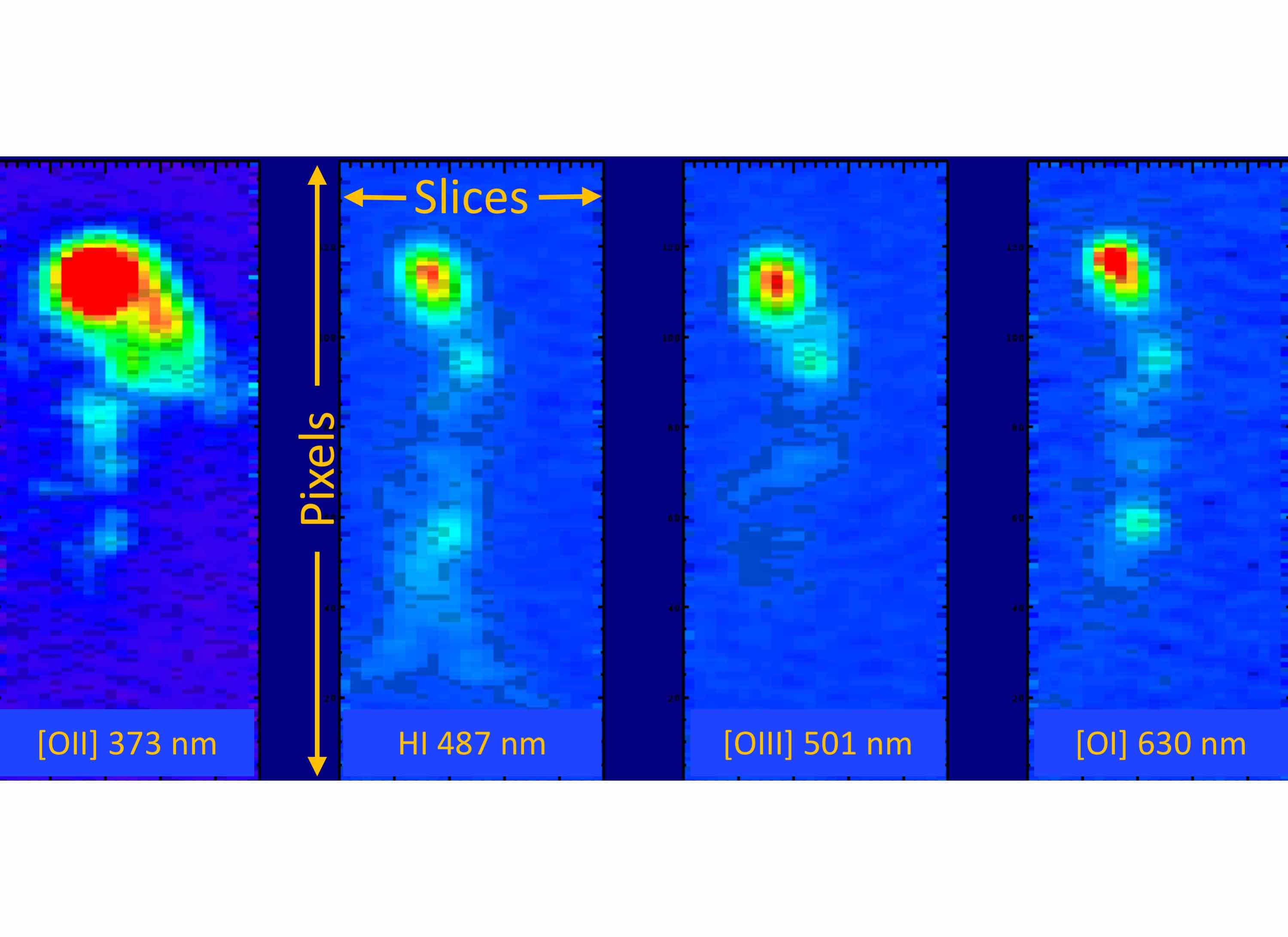}
\caption{Emission line reconstructions of HH32 extracted from the pipeline-produced -icubes files.  Pixels in the reconstruction are not square in order to show the image with the correct aspect ratio.\label{fig:drp_slices}}
\end{figure*}

Figure~\ref{fig:drp_slices} shows four reconstructed HH32 emission line images from the pipeline generated {\tt  $\_$icubes} files based partly on the data shown in Figure~\ref{fig:drp_images}, part of a series of spectra with a different grating configurations from the same night.  Since the cube contains several different atomic emission lines it can be used to reconstruct the emission of different gasses, or alternately the kinematic distribution of gas from one species.

\begin{deluxetable*}{cll}
\tablecaption{KCWI DRP Stages\label{tab_drp_stages}}
\tablehead{\colhead{Stage} & \colhead{Program} & \colhead{Description}}

\startdata
0	& kcwi\_prep		& Association of science and calibration files \\
1	& kcwi\_stage1		& Basic CCD-level reduction, nod-and-shuffle
panel subtraction \\
2	& kcwi\_stage2dark	& Master dark frame creation and subtraction \\
3	& kcwi\_stage3geom	& Geometry solution and creation of pixel maps \\
4	& kcwi\_stage4flat	& Illumination correction using b-splines \\
5	& kcwi\_stage5sky	& Sky subtraction using b-splines \\
6	& kcwi\_stage6cube	& Image cube creation \\
7	& kcwi\_stage7dar	& Differential atmospheric refraction
correction \\
8	& kcwi\_stage8std	& Standard star measurement and flux
calibration \\
\enddata
\end{deluxetable*}

\begin{deluxetable*}{clll}
\tablecaption{KCWI DRP Primary Output Products\label{tab_drp_products}}
\tablehead{\colhead{Stage} & \colhead{File Suffix} &
\colhead{Units} & \colhead{Notes}}

\startdata
1	& int	&	electrons		& reduced output intensity \\
	& var	&	electrons$^2$	& intensity variance \\
	& msk	&	flag		& cosmic rays, defects flagged \\
	& sky	&	electrons		& sky panel for nod-and-shuffle \\
	& obj	&	electrons		& object panel for nod-and-shuffle \\
\hline
2	& intd	&	electrons		& dark current subtracted intensity \\
	& vard	&	electrons$^2$	& variance + dark image variance \\
	& mskd	&	flag		& combination of dark and object flags \\
\hline
3	& geom	&	struct		& IDL structure with geometry solution \\
	& wavemap	&	Angstroms		& Map of wavelengths at each pixel \\
	& posmap	&	pixel		& Map of slice position at each pixel \\
	& slicemap	&	slice	& Map of slice number at each pixel \\
\hline
4	& intf	&	electrons		& flat-fielded intensity \\
	& varf	&	electrons$^2$	& variance + master flat variance \\
	& mskf	&	flag		& no change \\
\hline
5	& intk	&	electrons		& sky subtracted intensity \\
	& vark	&	electrons$^2$	& variance + master sky variance \\
	& mskk	&	flag		& no change \\
\hline
6	& icube	&	electrons		& 3D intensity data cube \\
	& vcube	&	electrons$^2$	& 3D variance cube \\
	& mcube &	flag		& 3D mask cube (padded pixels masked) \\
	& scube &	electrons		& 3D sky panel intensity cube \\
	& ocube	&	electrons		& 3D object panel intensity cube \\
\hline
7	& icubed &	electrons		& DAR corrected 3D data cube \\
	& vcubed &	electrons$^2$	& DAR corrected 3D variance cube \\
	& mcubed &	flag		& DAR corrected 3D flag cube (padded pixels masked) \\
\hline
8	& icubes &	$10^{16}$ ergs-s$^{-1}$-cm$^{-2}$-\AA$^{-1}$		& Flux: $F_{\lambda} \times 10^{16}$ \\
	& vcubes &	$(F_{\lambda} \times 10^{16})^2$	& Flux variance  \\
	& mcubes &	flag		& no change \\
\enddata
\end{deluxetable*}

\subsection{Pipeline Preparation}

Stage 0 of the pipeline reads all of the image headers from a given night to associate the
instrument configurations and their calibration sets.  Two files
are written that determine the way in which subsequent stages are
executed: {\tt  kcwi.ppar}, which sets pipeline parameters, and {\tt  kcwi.proc}, which lists the default
associations of object files with their required calibration files.
Additional files written out in this step include the
pipeline parameter files required to reduce master calibration images, and
an image summary file ({\tt  kcwi.imlog}).

\subsection{Basic CCD Reduction}

Stage 1 performs detector-specific processing such as bias subtraction, overscan trimming, gain correction, cosmic ray rejection, cosmetic defect
correction, error image creation, and scattered light
subtraction.  If the image was taken in nod-and-shuffle mode, this stage
also handles the re-packaging and subtraction of the sky panel from the
object panel.  The output of this stage is a two-dimensional image with
intensities in electrons.

Stage 2 creates and subtracts a master dark frame and then
scales and subtracts it from all object frames whose CCD configuration
matches the master dark.

\subsection{Geometry Solution and Map Generation}

Stage 3 solves the geometry of the three-dimensional data cube using continuum
bars and arc lamp images to define both the spatial and the spectral
mapping (see Figure~\ref{fig:drp_images}).  The bar target is projected across the slices in continuum light such
that each one has 5 bars that define the spatial scale along the slit height.
The arc lamp spectra are solved in
wavelength by comparing with an atlas (either FeAr or ThAr).
These wavelength solutions, combined with the known spatial separation of
the bars, produces the mapping required to assemble the three-dimensional
data cube.

The pipeline outputs the coefficients of the transformation for each slice
along with ancillary geometry information into a FITS table that will be used
to generate 3D data cubes in stage 6.  It also outputs three maps that give
the 3D data cube position for each 2D pixel: a slice map, corresponding to
the X axis in the 3D cube, a slice position map, corresponding to the Y axis,
and a wavelength map, corresponding to the Z axis.

\subsection{Illumination Correction}

Stage 4 corrects the profile of each slice image that results from vignetting, pixel response, and other factors.

Continuum flats are used to generate a master correction image that removes
the illumination and pixel-to-pixel variations in one step.  A set of six
continuum flat images are stacked and averaged with outlier rejection to
reduce the effect of cosmic rays.  A reference region on a single slice is
used to normalize the illumination of all other slices.  This reference region is fit
using b-splines with sufficient tension to detect pixel-to-pixel variations
within each individual slice.

The output of this stage is a master flat image for each configuration and
the illumination corrected images for all continuum flat and object images.

\subsection{Sky Subtraction}

Stage 5 uses the illumination corrected images, along with the geometry
maps produced in stage 3 to produce a model of the sky spectrum
using b-splines.  Object flux is rejected with a statistical cutoff.
For compact objects, this sky
subtraction is effective.
If the object fills the IFU, then a separate
sky observation should be used with appropriate scaling to account for sky
brightness variations.  The nod-and-shuffle technique is also available for the most critical background subtraction applications. 

Stage 5 produces a sky model 2D image and the sky-subtracted
2D object image.

\subsection{Data Cube Generation}

Stage 6 generates the 3D data cube.  This stage
uses the geometry mapping previously determined in stage 3 to transform each 2D
image into a geometry-corrected 3D data cube that reconstructs the focal plane at each wavelength bin.

\subsection{Differential Atmospheric Refraction Correction}

Stage 7 corrects differential refraction.  The atmosphere refracts light as a function of wavelength with the
result that at high airmass objects are pre-dispersed on the image slicer.  This has the effect of tilting spectra in the raw images, which can be accounted for in the data cubes.  Using the known properties of the atmosphere combined with the airmass, wavelength band, and orientation of
the input observation, the spatial shift at each wavelength can be
calculated.  Each wavelength slice of the data cube is shifted with a
fractional pixel shift to remove the differential shifts due to the
atmosphere.

\subsection{Flux Calibration}

Stage 8 uses standard
star observations to measure the instrumental response and scale science
observation to flux units.  Reference standard star spectra are from the
CALSPEC\footnote{http://www.stsci.edu/hst/observatory/crds/calspec.html} database hosted by the Space Telescope Science Institute (Baltimore, MD).  The light from the observed standard, which is usually the brightest source in the field, is integrated.
The reference spectrum is resampled onto the observed
wavelength scale and compared with the observed spectrum to produce an
inverse-sensitivity curve that is used to scale engineering
units to flux-calibrated units.  The
calibrated cubes are scaled by $10^{16}$ and give the brightness units in the header
as FLAM16.

\section{System Verification}
\label{s:performance}

\subsection{Laboratory Calibration}

\begin{figure*}
\plotone{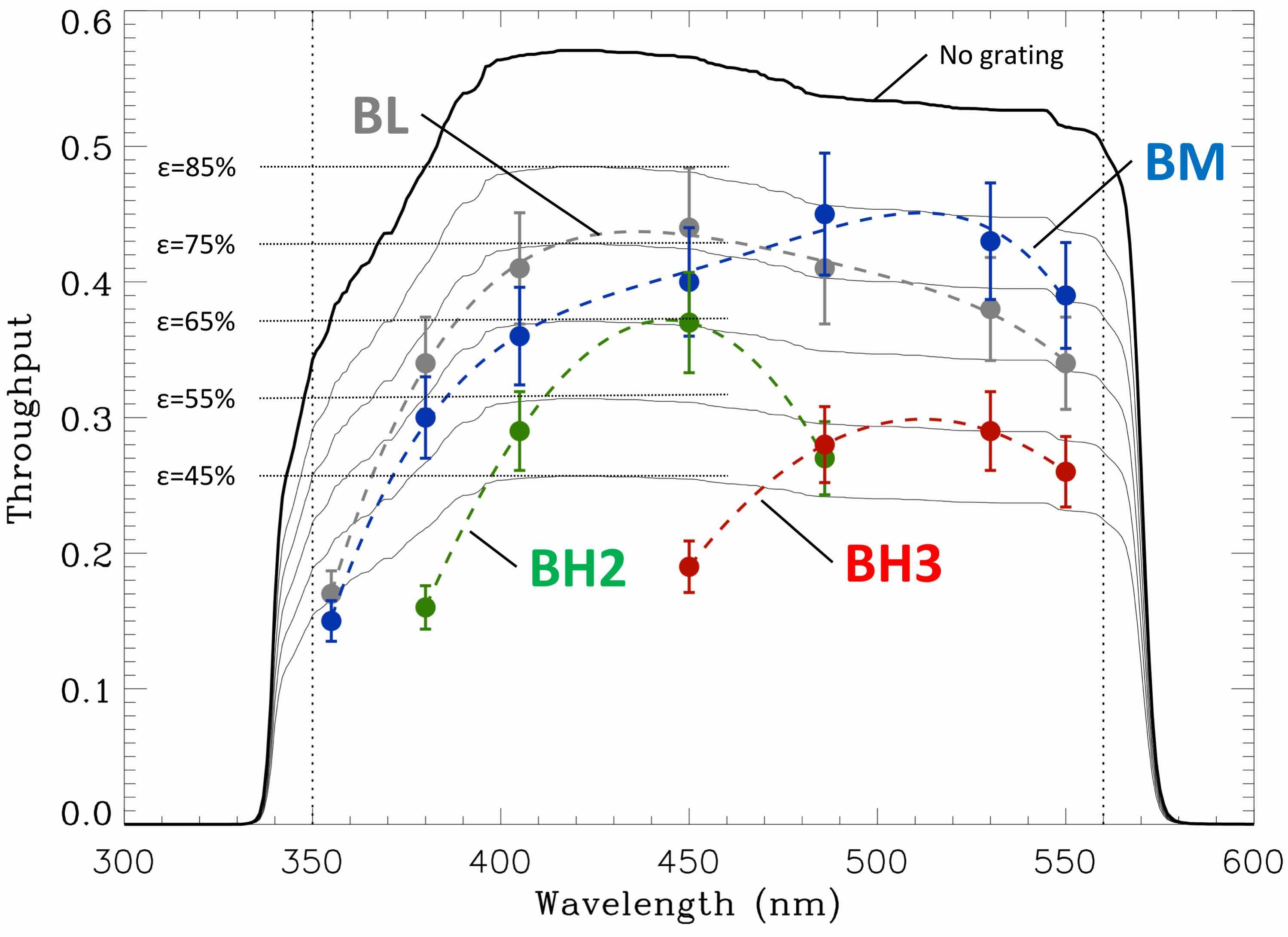}
\caption{The lab-measured end-to-end KCWI throughput is shown with filled circles for large slicer configurations shown in the indicated colors.  Polynomial fits to the laboratory data are over-plotted with dashed lines.  The model instrument throughput without a grating (but with the CCD) is shown with a solid black line, while a family of curves are scaled in light gray representing the system efficiency that would be achieved with grating throughput ($\epsilon$) as labelled.\label{f:labthroughput}}
\end{figure*}

The end-to-end optical performance of the fully assembled instrument was measured at $25 ^{\circ}$C in the laboratory at Caltech prior to shipment.  We projected a 0.5~mm diameter pinhole image onto the focal plane using the calibration system. A series of 10~nm full width at half maximum (FWHM) filters centered at 380~nm, 405~nm, 450~nm, 486~nm, and 530~nm defined the bandpass for five measurements in each grating configuration.  Each was referenced to the KCWI focal plane camera CCD sensor using the quantum efficiency from the manufacturer's data sheet.  The focal plane camera read noise (16~electrons) and gain ($1.26$~electrons-DN$^{-1}$) were measured with the photon transfer technique (\citet{janesick87}).  The large slicer was then inserted into the focal plane position (spot centered) with the grating rotation optimized for the wavelength under test.  Results are plotted in Figure~\ref{f:labthroughput} using filled circles.  They are corrected (reduced by a few percent) for the throughput of the guider dichroic, K-mirror assembly and window since these elements were not included in the beam path for our lab measurements.  We estimate the uncertainty to be about 10\%, including errors in the focal plane and science camera calibrations, variations in the LED source intensity, and scattering in the bandpass filters.  We show the model system efficiency as a solid line, which is the product of the instrument throughput (Figure~\ref{f:instnograting}) and the science camera QE (Figure~\ref{f:qe}).  We have over-plotted a family of curves with 45\% to 85\% grating throughput to indicate the grating performance implied by the end-to-end measurements for each configuration as a function of wavelength.  The implied peak throughput of the BL, BM and BH2 gratings, which includes the gelatin diffraction efficiency, absorption, and any reflection losses, meets our expectations while the BH3 grating has a peak efficiency around 55\% and could likely be improved.  Absorption in the grating gelatin and bonding agent probably accounts for a rolloff below 400~nm.  The reason for the reduced BH3 performance is not clear, but the most likely explanations are some combination of a misalignment in the peak for each polarization and scattering in the gelatin.  This grating has a slightly hazy appearance compared to the others and the vendor is working on a replacement to be delivered with BH1.

\begin{figure*}
\plotone{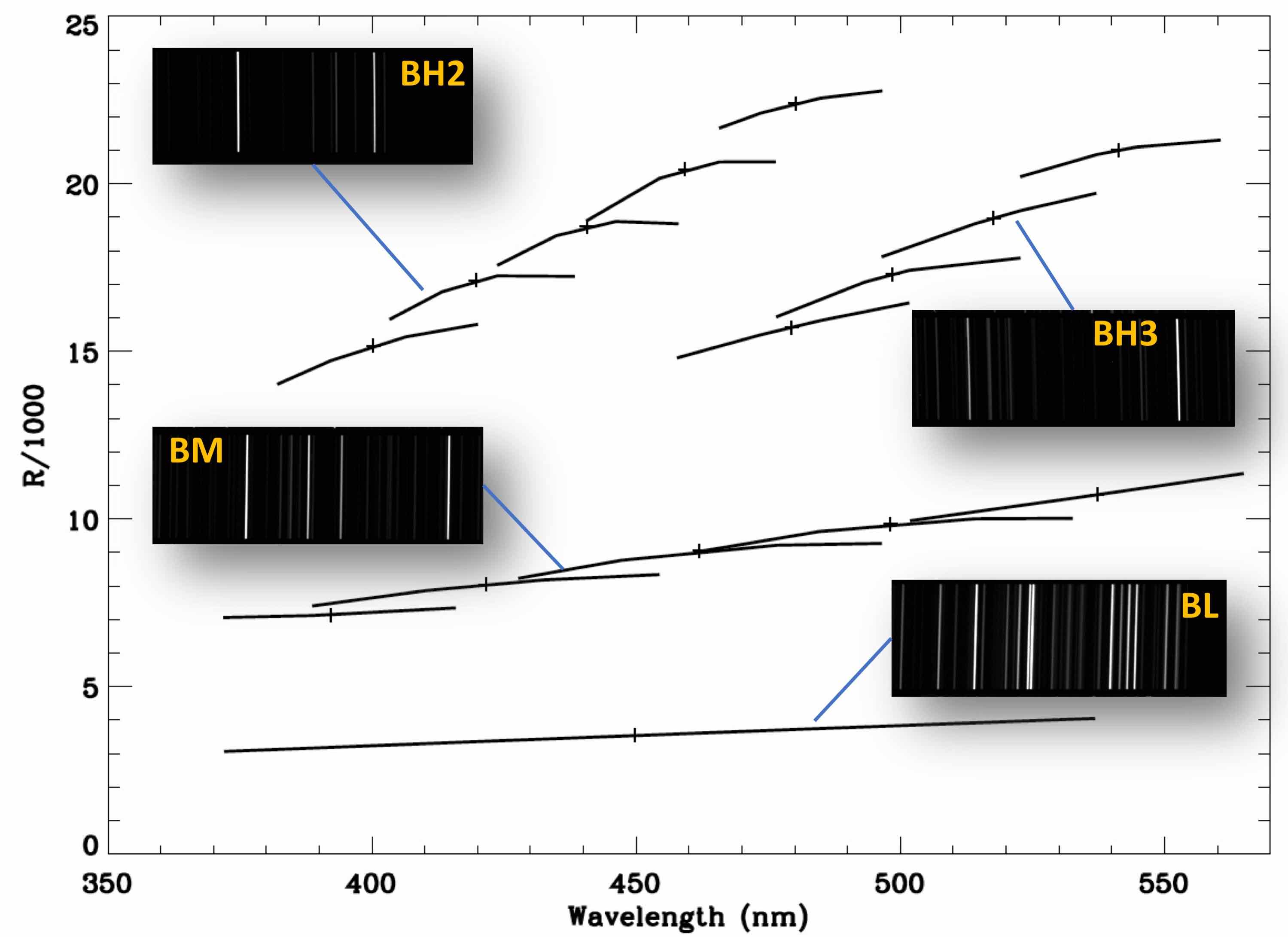}
\caption{Laboratory-measured spectral resolution of KCWI using the small slicer and each of the available gratings.  Points marked with a `+' indicate wavelengths for which the camera and grating were configured.  The inset calibration arc images were taken during commissioning using the small slicer and show sample imaging performance for each grating configuration.\label{f:labspecres}}
\end{figure*}

We also measured spectral resolution with the calibration unit FeAr hollow cathode lamp at Caltech prior to delivery. The instrument was configured with the small slicer and each of the four available
gratings.  The camera and grating rotations were configured optimally to span the efficient range of each grating.  The data was processed in ``light bucket'' mode, for which a single spectrum is generated for the entire slicer.  We subsequently evaluated the widths of isolated arc lamp lines to compute spectral resolution. The results are presented in Figure~\ref{f:labspecres} and demonstrate the widely configurable range of KCWI with spectral resolution achieved in excess of 20000.  The spectral resolution varies with wavelength, increasing as expected with the tangent of the grating angle.  Example small slicer slit image quality (similar for all gratings) is shown with $1 \times 1$ detector binning for samples of calibration arc data taken in operating conditions at Keck observatory.  The measured resolution is slit-limited for all medium and large slicer configurations, which would scale the results in the Figure by factors of 0.5 and 0.25 respectively.

\subsection{Commissioning}

KCWI was commissioned on the Keck II telescope during three observing runs over ten nights in 2017 as shown in Table~\ref{tbl:schedule}.  Only two of the nights were mainly clear, but many of the required tasks (pointing, de-rotator alignment, N\&S testing) could be adequately evaluated with relatively bright stars.  During the program, the observatory also commissioned its telescope control system upgrade (\citet{kwok16}) for KCWI and an immediate improvement in the blind pointing precision from $\sim 30\arcsec$ to $\sim 3\arcsec$ was noted, well within the required precision to support nod and shuffle observations even with the $8.25\arcsec \times 20\arcsec$ small slicer.  In this section, the key performance verifications for spatial and spectral resolution are described.

\begin{deluxetable}{rcl}
\tablecaption{Commissioning Conditions and Schedule\label{tbl:schedule}}
\tablehead{
\colhead{Date} & \colhead{Lunar Phase} & \colhead{Conditions}}
\startdata
11 April       & 15d              & Partly cloudy \\
13 April       & 17d              & Partly cloudy \\
14 April       & 18d              & Partly cloudy \\
13 May         & 17d              & Clear \\
15 May         & 19d              & Cloudy \\
16 May         & 20d              & Partly cloudy \\
15 June        & 20d              & Partly cloudy \\
18 June        & 23d              & Partly cloudy\\
19 June        & 24d              & Partly cloudy \\
20 June        & 25d              & Clear\\
\enddata
\end{deluxetable}

Throughput was measured for the white dwarf standards listed in Table~\ref{tbl:standards}, all of which are included in the CALSPEC database (e.g. \citet{turnshek90, bohlin14}).   We measured the standards periodically over the course of commissioning with a range of grating, filter and slicer combinations, although a large fraction are from the night of 15 June.  

\begin{deluxetable}{crrr}
\tablecaption{KCWI CALSPEC Standards\label{tbl:standards}}
\tablehead{
\colhead{Star} & \colhead{m$_V$} & \colhead{R.A.} & \colhead{Dec.}}
\startdata
HZ4        & 14.51 & 03 55 21.990  & +09 47 18.00 \\
G191B2B    & 11.78 & 05 05 30.613 & +52 49 51.96 \\
Feige67    & 11.82 & 12 41 51.791 & +17 31 19.76 \\
HZ44       & 11.67 & 13 23 35.258  & +36 07 59.51 \\
BD+33 2642 & 10.83 & 15 51 59.886 & +32 56 54.33 \\
BD+02 3375 &  9.93 & 17 39 45.596 & +02 24 59.60 \\
BD+28 4211 & 10.51 & 21 51 11.021 & +28 51 50.36 \\
BD+25 4655 &  9.69 & 21 59 41.975 & +26 25 57.40 \\
\enddata
\end{deluxetable}

The analysis was performed on bias-subtracted KCWI raw data, using the pipeline-produced ``wavemap'' files (created from sets of associated calibration lamp spectra) to identify the wavelength associated with each pixel.  The selected standards are the dominant sources in each field.  In order to estimate the signal, the columns of each image were sorted according to total integrated brightness including all wavelengths.  The brightest 10\% of columns (approximately equivalent to two complete slices) were integrated to form the target signal, while the next 10\% of columns were integrated to form the sky signal, thus ensuring that the sky was measured in the vicinity of the source.  We also estimated the fractional contribution of each sky measurement and rejected those with a contribution greater than 5\%.  The typical sky brightness measured in this way ranges from $1-3$\% of the signal.

\begin{figure*}
\plotone{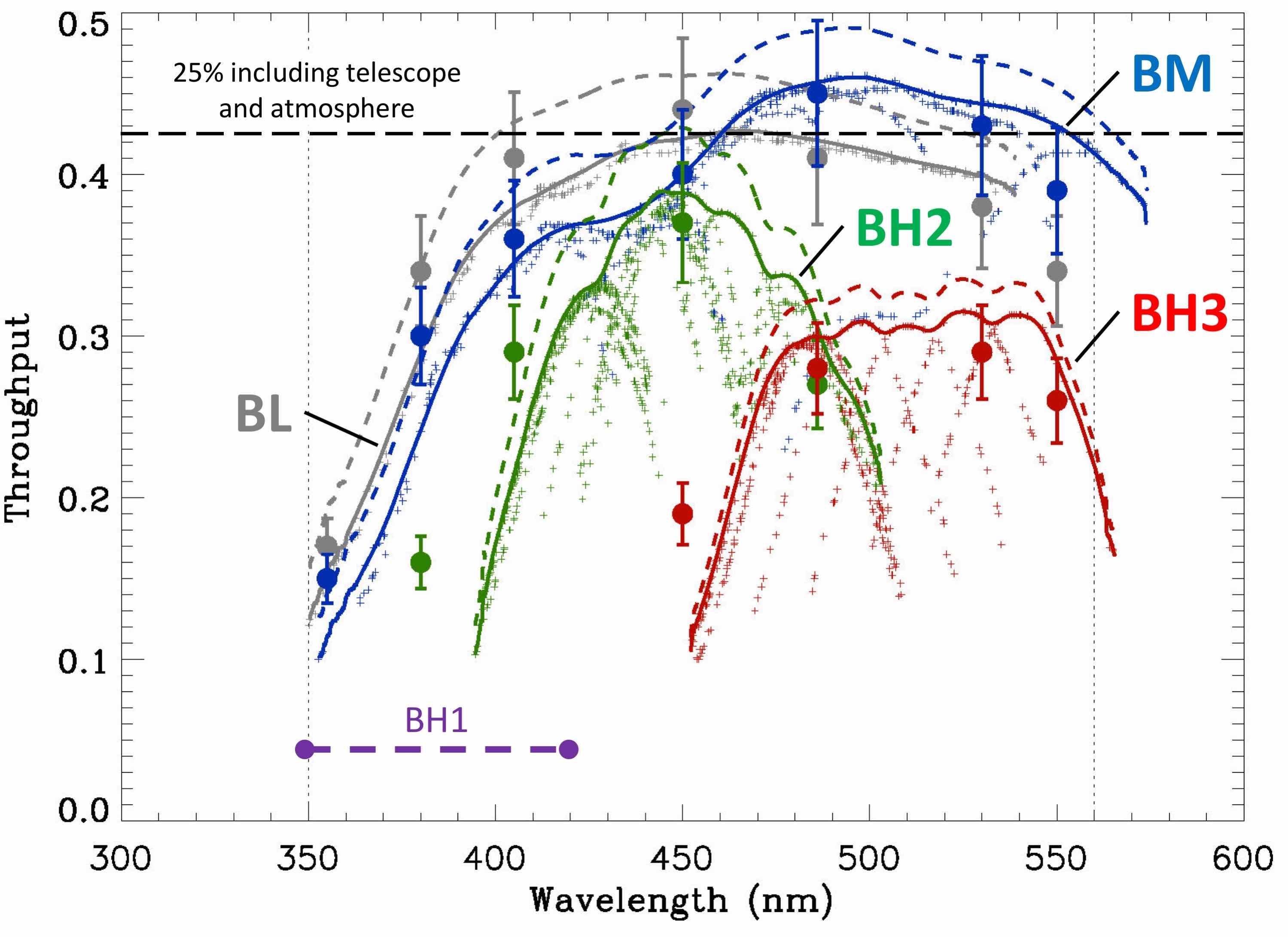}
\caption{A composite of CALSPEC standard star throughput measurements for the large slicer and four available KCWI gratings, corrected for the telescope and atmosphere.  Each grating is shown with a different color, including a placeholder for the BH1 grating.  Individual configuration data is shown with $+$ symbols, while the envelope of measurements for all on-sky configurations for a given grating is shown with a solid line.  In order to indicate the uncertainty in the measurements, a colored dashed line is also shown with extinction $1.5\times$ worse than photometric.  Lab-measured data points are shown with filled circles.  The black dashed line indicates the throughput that would be measured with the telescope and atmosphere included.
\label{f:instthroughput}}
\end{figure*}

The end-to-end throughput of KCWI corrected for telescope and atmosphere for 74 large slicer spectra is presented in Figure~\ref{f:instthroughput}.  Measurements for each configuration are shown with $+$ symbols, while the envelope of each set of measurements for a given grating is shown with a thick solid line.  Colors are used to identify measurements for each of the four gratings, including an indication of the design band of the BH1 grating currently in fabrication.  Pre-ship laboratory measurements of the throughput from Figure~\ref{f:labthroughput} are overplotted with filled circles.  The raw throughput measured with telescope and atmosphere included is shown with a black dashed line for scale.  The losses due to the reflectivity of the three aluminum-coated Keck telescope mirrors can not be independently measured.  We have assumed a measured reflectivity curve for fresh coatings provided by the observatory (typically about $90$\% in the KCWI band) with approximately $1.5$\% reflectivity loss for each Keck mirror based on the average 1 year age of the coatings at the time of the KCWI commissioning (personal communication Kuochou Tai).  We have also made a correction for atmospheric extinction based on \citet{buton13} and the airmass of the standard star at the time of the observation.  Since we do not have a large number of measurements on different days, and since the weather was not perfect, we also show the instrument response that would be derived under less than ideal conditions as an estimate for the uncertainty with solid dashed lines.  These latter models use the Buton extinction scaled by a factor of 1.5.

The agreement between the pre-ship end-to-end measurements and the on-sky measurements is excellent, especially given the uncertainties.  Measurements with the medium slicer are similar, and measurements with the small slicer are a few percent lower.  On a bright note, the coating program for KCWI was quite successful and the complete system exhibits end-to-end throughput $1.5-2.5\times$ higher than PCWI before accounting for the large increase in telescope aperture.  With telescope and atmosphere included, the Keck II+KCWI system achieves a peak throughput of over 25\%.


\subsection{Spectral Resolution}

\begin{figure*}
\plotone{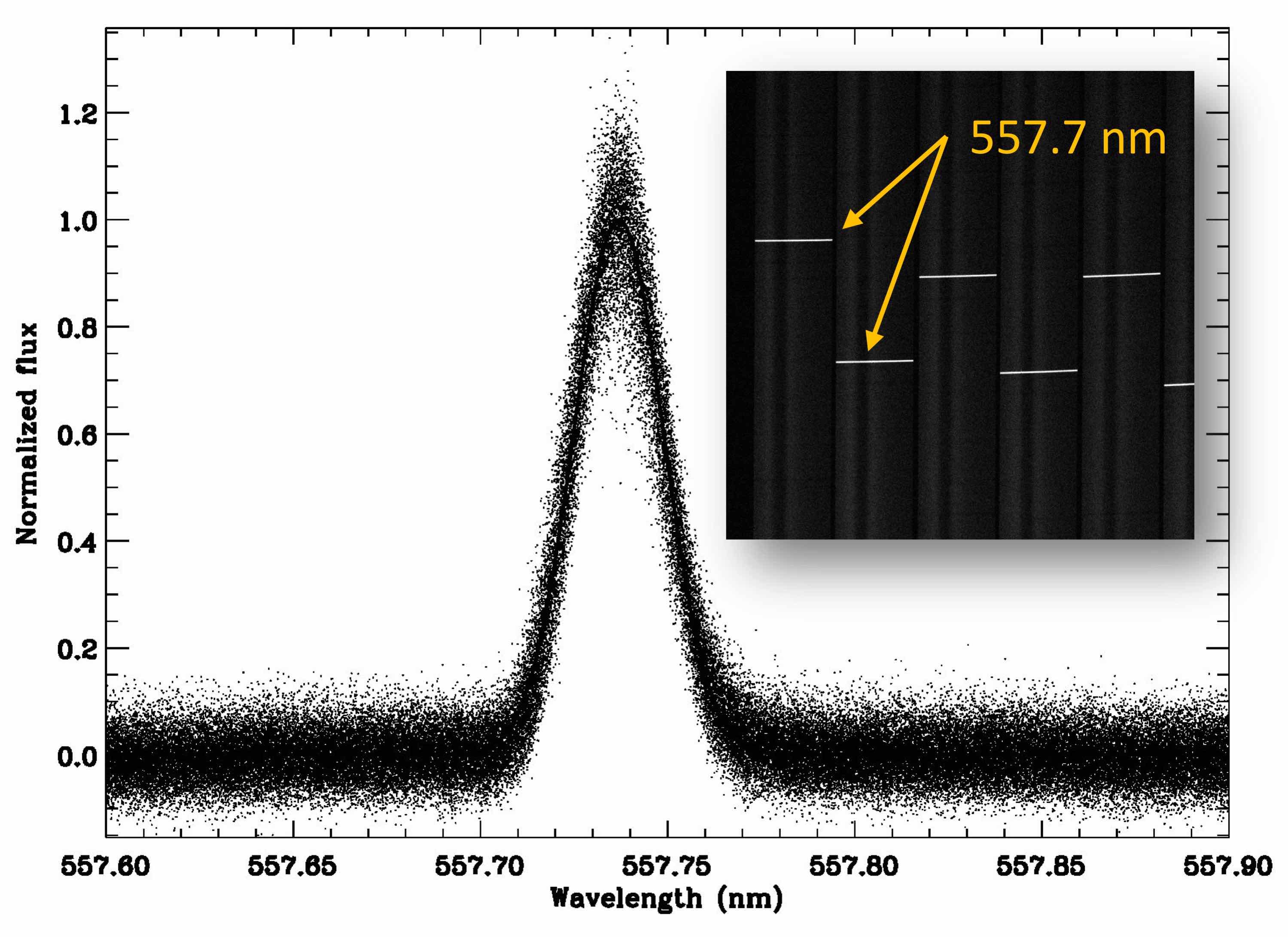}
\caption{A light bucket mode sky emission line extraction showing on-sky spectral resolution close to 20000 with the small slicer and the BH3 grating.  The inset shows the bright sky line at 557.7~nm used to generate the curve.\label{f:skyspecres}}
\end{figure*}

We re-measured the spectral resolution on sky during the first commissioning run. We observed the galaxy NGC5886 for 3 minutes using the small slicer and BH3 grating centered at 545.5 nm. In this setting, we capture the very
bright [O I] atmospheric line at 557.7 nm.  We analyzed the line the same way as
the arc spectra in lab (which provides a dithered spectral line measurement as the different slices image the line with different phasing across the detector). We found a spectral resolution R$\sim$19,500, comparable but slightly below the pre-ship value. Figure~\ref{f:skyspecres} shows the ``light bucket mode'' sky line
and our fit to it.  We also observed a $\sim 5$\% variation in spectral resolution across the 24 slices primarily due to residual focus non-uniformity across the CCD and slit rotation effects.  The system is only resolution-limited for the high resolution gratings and small slicer.  Other configurations are slit-limited.


\section{Observing Examples}
\label{s:examples}

In this section we present a sample of observations made with KCWI during commissioning.  The examples are selected to show a range of KCWI applications and to familiarize the reader with KCWI data and pipeline reconstructions for likely categories of targets.

\subsection{First light image of M3 with KCWI} 

\begin{figure*}
\plotone{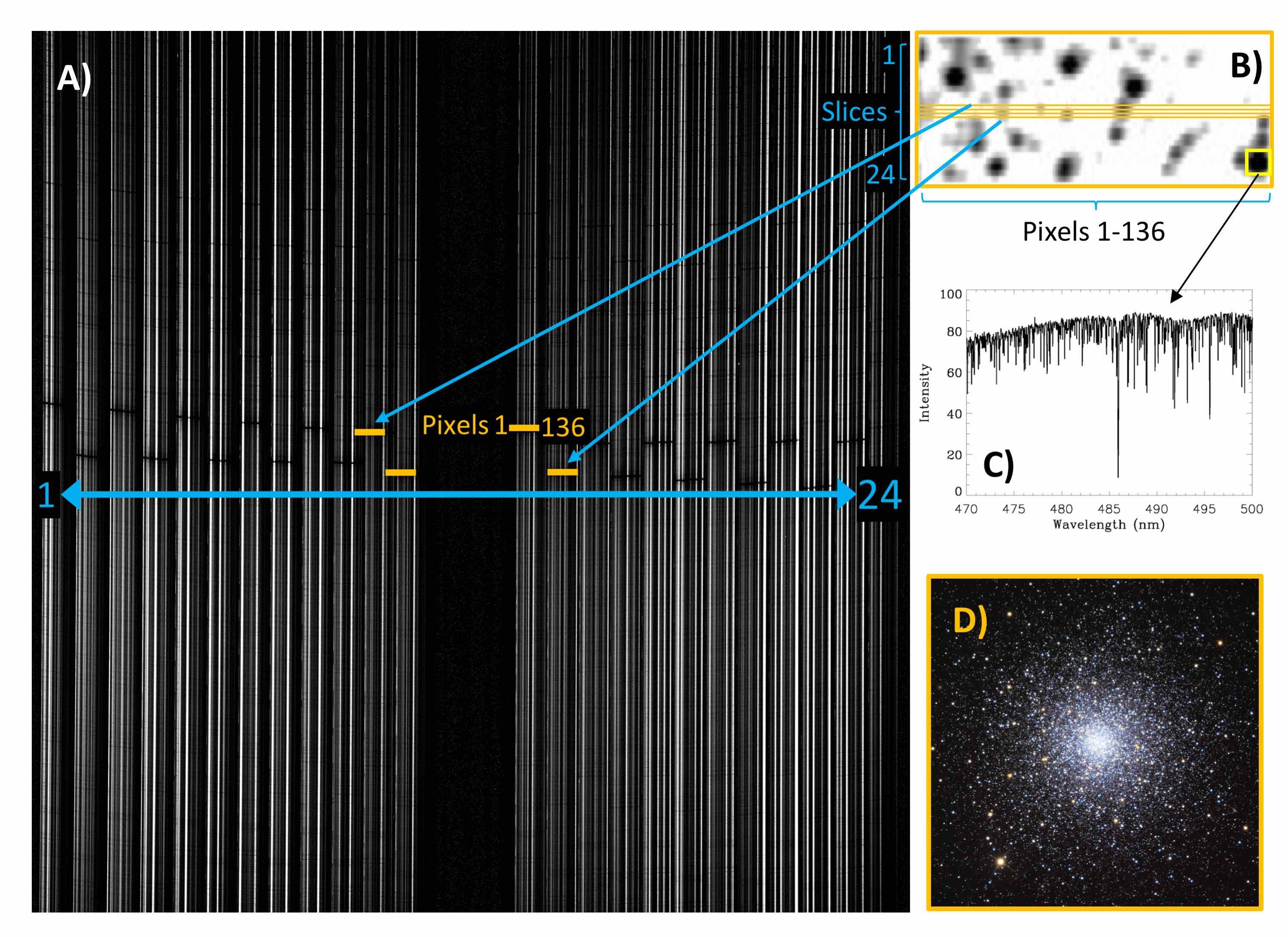}
\caption{KCWI first light image of M3.  Panel A shows the pipeline-reduced $\_$intf spectral image.  Panel B shows the reconstructed $20\arcsec \times 8.25\arcsec$ image of the cluster core.  Panel C shows a representative R$\sim 20000$ stellar spectrum, and panel D shows the full $18'$ extent of the cluster (credit for M3 image: Adam Block/Mount Lemmon SkyCenter/University of Arizona)
\label{f:cmartinf1}}
\end{figure*}

Our first example illustrates the high spatial and spectral resolution capability of KCWI pairing the small slicer with the BH2 grating. In Panel A of Figure~\ref{f:cmartinf1}, we show a spectrum  of the core of the globular cluster Messier 3 centered near the H$\beta$ line at 486.1~nm. Panel B in the upper right shows the reconstructed image on the small slicer.  The reconstructed image is oriented so that the 24 slice images are horizontal (the same as in the Panel A).  Blue arrows indicate how slice images are aligned to form representative parts of the image.  Full spectral information is contained in the reconstructed image cube.  A rich example spectrum of one star with many absorption lines is shown in panel C.  A wide field image (courtesy Adam Block/Mount Lemmon SkyCenter/University of Arizona) showing the full $18'$ extent of the cluster is shown in panel D.  The high spectral resolution (R$\sim 20000$) allows us to measure star velocities with a precision of $\leq 10$ km-s$^{-1}$) and high sensitivity to many elemental spectral lines. This data cube was obtained with a single 10 minute exposure, with a total field size of $20\arcsec \times 8.3\arcsec$. The seeing-limited stellar images are approximately $0.7\arcsec$ wide.

\subsection{Feedback from Star Formation in a Spiral Arm} 

\begin{figure*}
\plotone{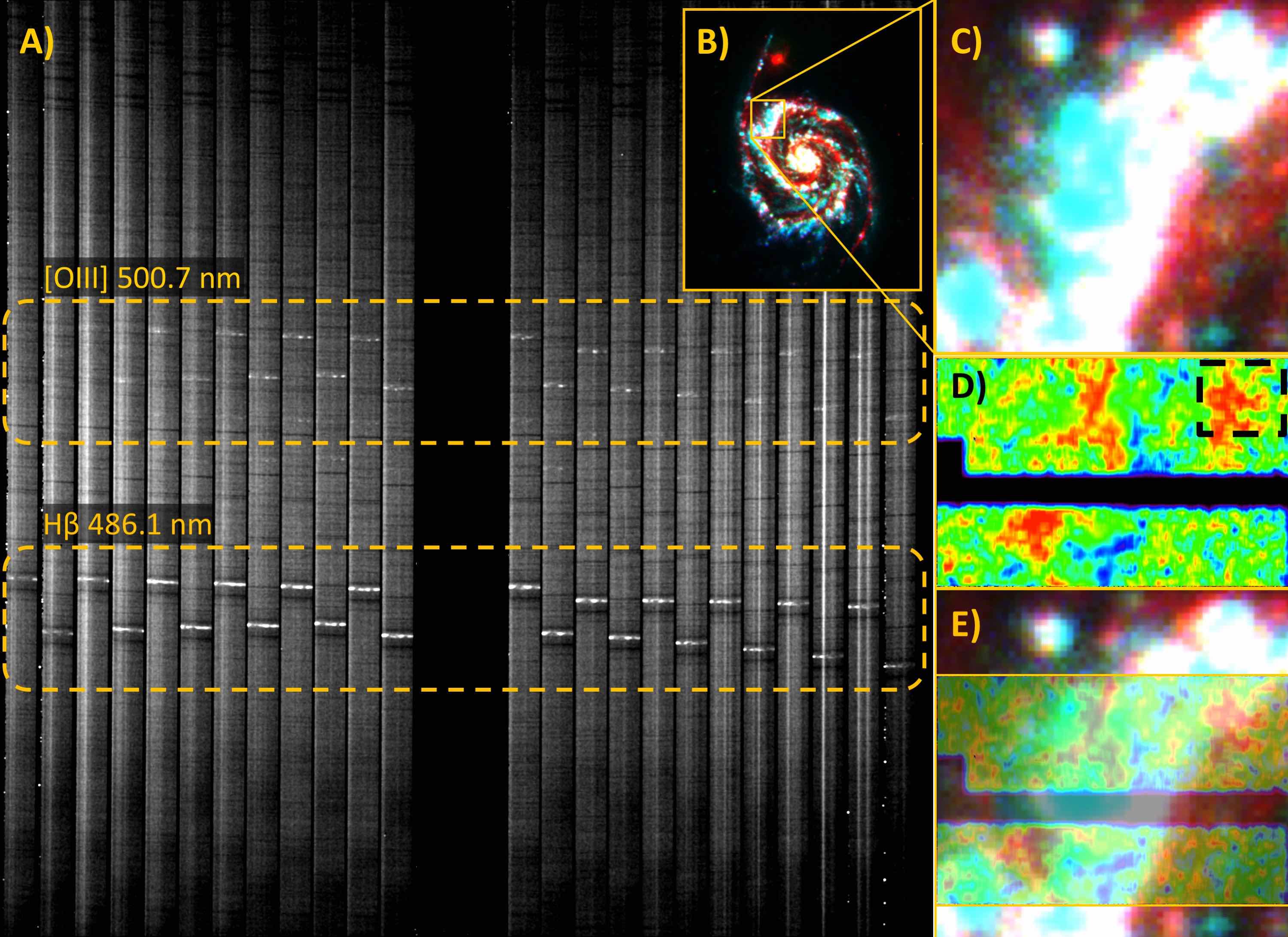}
\caption{KCWI reveals the direct connection between feedback from massive star formation and the regulation of that star formation.  Panel A: A 300s KCWI spectrum of M51 with the medium slicer and BH3 grating.  Panel B: A GALEX+Spitzer image of the Whirlpool Galaxy M51.  Panel C: A blowup of a spiral arm imaged by GALEX+Spitzer that is the subject of the current KCWI study.  Panel D: A mosaic of KCWI [OIII] 500.7~nm images of M51.  The footprint of the medium slicer is shown with a black dashed rectangle.  Panel E: The KCWI mosaic overlaid on the GALEX+Spitzer image showing excellent correlation between the KCWI emission and the GALEX UV emission.\label{f:cmartinf2}}
\end{figure*}

One of the most important problems in understanding star formation and galaxy evolution is how feedback from massive stars regulates star formation. KCWI can directly measure gas that has been shocked by massive star winds and supernova explosions. In the image montage shown in Figure~\ref{f:cmartinf2}, we focus on a spiral arm of the Whirlpool Galaxy, Messier 51. The spiral arm is created by a spiral density wave that triggers a burst of star formation.  

This example shows a mosaic of 300~s images collected with the medium slicer and the high dispersion BH3 grating demonstrating excellent spatial ($\sim 0.7$\arcsec) and spectral (R$\sim10000$) resolution.  The set of observations is comprised of 11 tiles.  An example tile is shown in Panel A of Figure~\ref{f:cmartinf2}, in which the emission from H$\beta$ and OIII is identified.  In Panel B, we show a $7\arcmin \times 11\arcmin$ GALEX UV (blue) and Spitzer IR (red) overlaid image of M51.  A box superimposed on the image shows the region tiled by KCWI and is expanded in Panel C.  In this blowup, we can see the time sequence of events after the wave passes as stars move from lower right to upper left. Massive stars are traced by GALEX UV and gas and dust are traced by Spitzer IR. Star formation is triggered by the spiral wave shock creating a region of high density gas and dust and dust-obscured young stars (red = high IR to UV). The stars begin to feedback against the gas and dust allowing more UV to escape (yellow = intermediate IR to UV). Finally, the stars begin to explode in supernovae after 10-20 million years and clear out the dust and gas (blue = high UV to IR). 

In Panel D, we show a mosaic in the light of [OIII]~500.7~nm of the 11 KCWI pointings (a region approximately $68\arcsec \times 44\arcsec$ in size) with a gap that will be filled in at a later time.  KCWI traces the shocked, high velocity gas energized by these supernovae. Red regions are high velocity shocked gas. These are correlated with the UV bright regions as shown in Panel E, forming a web of shocked gas walls as the supernova shock waves push the gas away from the star forming regions and up and away from the galaxy's disk. As the gas is disrupted star formation stops (the arm peters out). This is the fundamental process which regulates star formation efficiency, and with this new KCWI result this process is now revealed and provides missing input to galaxy formation models.

\subsection{QSO-illuminated interacting disks}  

\begin{figure*}
\plotone{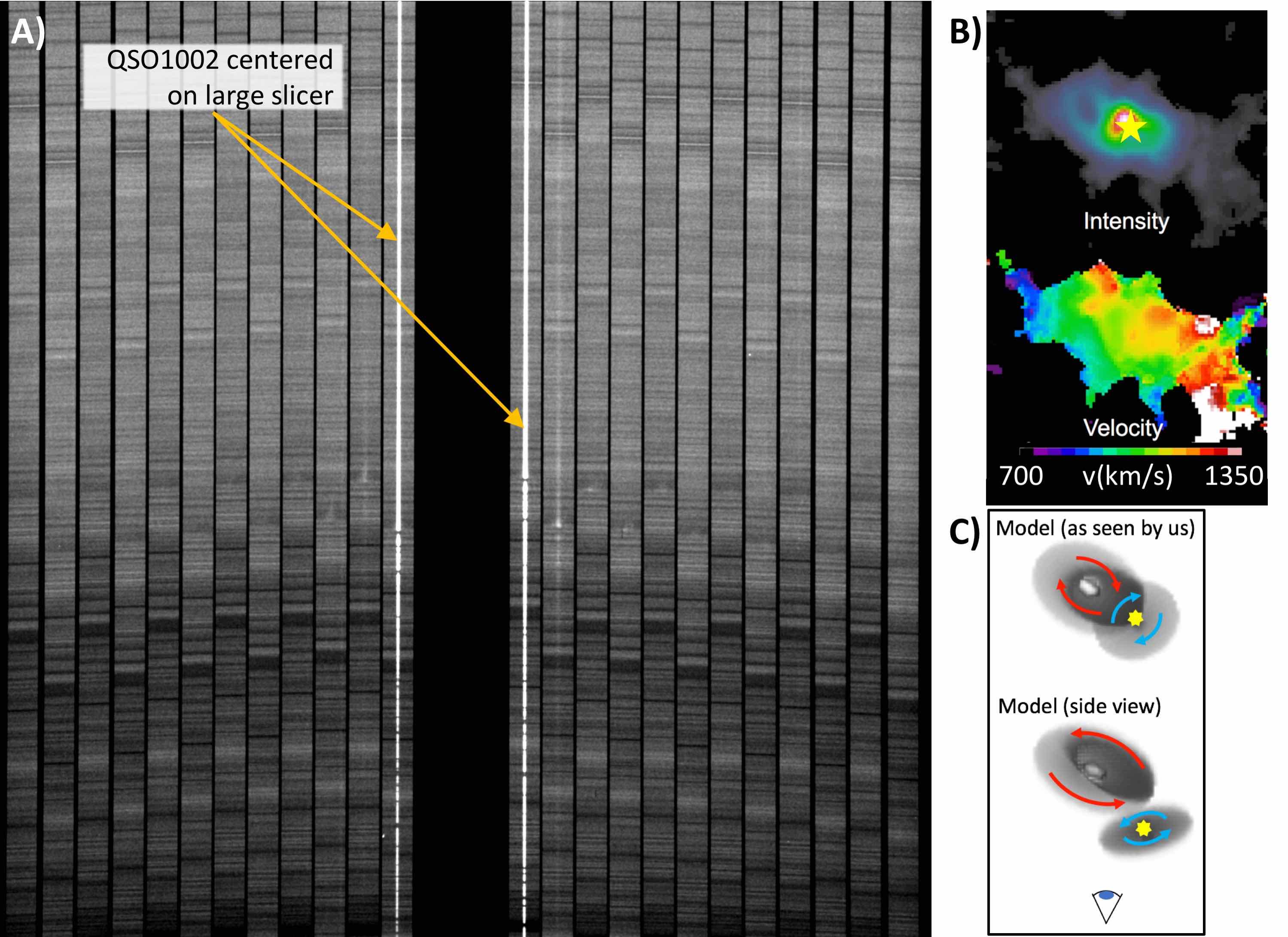}
\caption{Two giant disks illuminated by QSO1002.  Panel A shows a 1200~s $\_$intf spectral image with the bright quasar spectrum centered on the large slicer with the BM grating configured for the spectral band near 449~nm.  Panel B shows the reconstructed image on the slicer (top) and velocity map (bottom), which was determined by the velocity associated with the H$\beta$ 486.1~nm emission line.  Panel C shows a model that explains the emission as seen from our vantage point.\label{f:cmartinf3}}
\end{figure*}

Our third example shows the large slicer with BM grating to illustrate the low-surface brightness capability of KCWI. Two giant disks illuminated by quasi-stellar object QSO1002 are revealed in a single hour of exposure. KCWI observes the hydrogen H$\beta$ line emission from this distant object, and the redshift of the emission line probes the velocity of the gas in the system. A 20~m spectrum of the centered quasar and environs is shown in Panel A of Figure~\ref{f:cmartinf3}.  A narrow-band reconstructed intensity image (top, with QSO position marked by a yellow star) and gas velocity (bottom) are shown in Panel B of the Figure.  The bright QSO is model-subtracted and the background removed using a portion of the field known (from previous PCWI measurements) to be free from emission.
Red indicates gas moving away from us (redshifted), while blue/green indicates gas moving toward us. The image and velocity maps are consistent with the model shown in Panel C.
The model consists of two giant (60 kpc and 80 kpc) disks, rotating approximately with the same spin direction, and orbiting each other also with the same spin sense. The lower disk has the QSO in its center. The interaction of the two disks produces large torques and gas flows into the disk centers. The gas flow into the center of the lower disk is feeding the QSO. The light from the QSO exits in conical beams which only illuminate part of the companion disk (darker area of model and brighter area of image). KCWI can measure the disk size, rotation speed, gas inflow speed, and approximate separation. This detection reveals both the presence of large gaseous rotating disks in the distant universe and their role in generating the most luminous objects in the universe.

\section{Discussion}
\label{s:discussion}

KCWI has been successfully constructed and installed on Mauna Kea at the 10~m Keck II telescope.  The instrument is performing very well and is poised with its unique configurability and available nod and shuffle mode to open a new window to the low surface brightness universe.  Spectral resolution is excellent, and the spatial resolution is pixel-limited, taking full advantage of the excellent seeing available on the mountain.  The peak throughput is quite high for the BL, BM and BH2 gratings, and plans are underway to complete the grating suite with the BH1 installation and a BH3 upgrade.  The recent control system upgrade at the Observatory will enable the instrument to efficiently employ its nod and shuffle observing mode for the most challenging faint targets.

A conceptual design of the red arm was presented at the KCWI Preliminary Design Review in 2011.  Now that the phased delivery of the blue channel is successfully in place, work on the detailed design of the funded red channel is underway.  The new arm will include a large dichroic to replace the FMD fold mirror as well as an updated guider design that is compatible with red and blue science from $350 - 1050$~nm.  A second camera, detector and red-optimized grating suite are planned for the available volume on the top of the bench behind the FMD fold mirror.

\acknowledgments

The Keck Cosmic Web Imager was developed through a collaboration of the California Institute of Technology, the University of California, and the W.~M.~Keck Observatory.  The research described in this publication was carried out by the California Institute of Technology and by the Jet Propulsion Laboratory, which is managed by the California Institute of Technology.  The work was sponsored by grants from the the Telescope System Instrumentation Program (TSIP) and the Major Research Instrumentation Program (MRI) of the National Science Foundation, as well as grants from the Heising-Simons Foundation, the W. M. Keck Observatory, and the Caltech Division of Physics, Math and Astronomy.  We acknowledge our major industrial partner Winlight Optics (Pertuis, France) for their cooperation and assistance in the fabrication of the IFU and powered reflective optics.  We would also like to thank the referee for constructive and helpful comments.

Much of the data presented herein were obtained at the W. M. Keck Observatory, which is operated as a scientific partnership among the California Institute of Technology, the University of California and the National Aeronautics and Space Administration. The Observatory was made possible by the generous financial support of the W. M. Keck Foundation.

The authors wish to recognize and acknowledge the very significant cultural role and reverence that the summit of {\em Maunakea} has always had within the indigenous Hawaiian community.  We are most fortunate to have the opportunity to conduct observations from this mountain.

\facilities{GALEX, Hale, Keck:II (KCWI), Spitzer}

\end{document}